\documentclass[onecolumn,draftcls]{IEEEtran}
\usepackage{amsmath,amsfonts}
\usepackage{algorithmic}
\usepackage{algorithm}
\usepackage{array}
\usepackage[caption=false,font=normalsize,labelfont=sf,textfont=sf]{subfig}
\usepackage{textcomp}
\usepackage{stfloats}
\usepackage{url}
\usepackage{verbatim}
\usepackage{graphicx}
\usepackage{cite}
\usepackage{amssymb}

\usepackage{float}

\usepackage{cite}
\usepackage{xfrac}
\usepackage{wrapfig}
\usepackage[colorlinks=true, urlcolor=blue, linkcolor=blue, citecolor=blue]{hyperref}
\usepackage{tabularx,ragged2e,booktabs,caption}
    
\usepackage{xcolor}
\usepackage{url}

\newcommand{\rf}{\textcolor{black}}

\newcommand{\ta}{\textcolor{teal}} 

\usepackage{graphicx} 
\usepackage{float} 

 \usepackage{cmbright}

\begin{document}

\title{The ISAC systems aided by MIMO, RIS and with Beamforming Techniques}

\author{Rafael Augusto Pedriali, Isadora Martines Ferreira, Jose Carlos Marinello, and Taufik Abrão
\thanks{This work was supported in part by the National Council for Scientific and Technological Development (CNPq) of Brazil under Grants {314618/2023-6, 405301/2021-9}, and in part by CAPES under Grant Financial code 001 (numero contrato/processo pos-doc estratégico).}

\thanks{R. A. Pedriali is a post-doc at Lab. Telecom-UEL in the Electrical Engineering Department, State University of Londrina (UEL), Parana, Brazil}
\thanks{T. Abrão is an associate professor at the Electrical Engineering Department, State University of Londrina (UEL), PR, Brazil (e-mail: taufik@uel.br).}
}

\markboth{ISAC optimization}%
{Pedriali \MakeLowercase{\textit{et al.}}: Optimizing ISAC systems aided by MIMO, RIS
and with Beamforming Techniques}


\maketitle


\vspace{5mm}
\color{black}
\begin{abstract}
This paper explores the integration of communication and sensing in modern wireless systems through the configuration of BS and RIS antenna elements. By leveraging time multiplexing for both communication and sensing, the proposed system optimizes spectral efficiency and operational performance. The use of static RIS configurations tailored to specific environments eliminates the need for dynamic reconfigurations, enhancing system agility, reducing processing complexity, and improving sensing accuracy. The system incorporates trilateration, angle of arrival, and time of arrival techniques to enable precise user localization by combining signals reflected along multiple paths. This approach ensures a link selection and reduces the sensing costs while avoiding channel conflicts with communication data and emphasizing the importance of combining innovative technologies such as passive and adaptive beamforming in a unified framework.
\end{abstract}

\begin{IEEEkeywords}
ISAC, passive beamforming, sensing.
\end{IEEEkeywords}

\section{Introduction}
The demand for high-capacity, high-precision wireless networks in dense urban and indoor environments has driven new technologies to address limitations like signal attenuation, interference, and limited adaptability to dynamic channel conditions. The reconfigurable intelligent surfaces (RISs) have emerged as a passive solution, optimizing signal propagation and enhancing spectral efficiency \cite{RIS}. 
This technology enables dynamic control over signal propagation, enhancing channel conditions while mitigating issues like fading and interference. Combined with MIMO systems and adaptive beamforming at millimeter wave frequencies, RIS improves communication in both line-of-sight (LoS) and non-line-of-sight (NLoS) scenarios \cite{kazemi2019k, Beamformin2021}. Particularly in NLoS conditions, RIS creates optimized alternative paths, mitigating blockages. Additionally, RIS-assisted beamforming localization techniques increase accuracy, allowing precise determination of user and object positions in complex three-dimensional settings. These integrated sensing and communication (ISAC) systems significantly enhance overall efficiency through structured design \cite{ISAC1}.

However, LoS and NLoS conditions \cite{kazemi2019k} are common in wireless communication and significantly influence localization, particularly in urban and indoor environments where signals are often obstructed by buildings and other objects. In multipath scenarios, multiple reflections lead to interference and signal overlap, making it challenging to accurately determine the origin of the signal. Overcoming these challenges requires systems capable of efficiently and dynamically managing the radio channel under NLoS conditions, making solutions such as RIS \cite{RIS} and adaptive beamforming \cite{Beamformin2021} increasingly attractive. Beamforming manipulates the propagation direction of radio waves using an antenna array, while RIS dynamically adjusts the properties of waves reflecting on them \cite{ISAC1}. These adaptive solutions maximize signal quality and localization accuracy in complex scenarios. The implementation of these advanced technologies meets the immediate demands of dense urban environments and also stimulates the development of more robust and precise communication systems. These advancements support various applications, including intelligent transportation systems, emergency services, security operations, and the expanding drone and autonomous vehicle industries \cite{zhou2019successive}. In these domains, the capability to achieve accurate real-time radio mapping significantly enhances operational efficiency and safety.

The integration of ISAC into telecommunications systems enables platforms to acquire detailed information about their surrounding physical environment, facilitating simultaneous communication and sensing \cite{ISAC1}. This capability is particularly valuable in urban and indoor environments, where challenges such as signal reflections, obstructions, and multipath propagation are prevalent. This is achieved by ISAC-equipped devices capturing technical parameters such as time of arrival (ToA), angle of arrival (AoA), and received signal strength. These parameters enable the deduction of the layout and characteristics of the environment, even in indoor locations where GPS systems are less effective due to the absence of LoS with satellites \cite{lee1994efficient}.

In dense urban environments, managing the radio frequency (RF) spectrum is a persistent challenge due to high connectivity demands. ISAC systems address this issue by optimizing spectrum usage, combining communication and sensing within the same spectrum band \cite{ISAC1}. This approach reduces the need for separate bandwidth allocations for different functions. Additionally, the ability to provide real-time environmental data enhances operational decision-making, enabling fast and informed responses.

For instance, ISAC optimizes intelligent transportation systems operations by dynamically adjusting routes, monitoring traffic conditions in real time, identifying obstacles, and suggesting alternative paths for autonomous vehicles. Simultaneously, it issues timely alerts to drivers, improving safety and efficiency \cite{RIS}.
Additionally, the implementation of ISAC involves integrating radar and telecommunications sensors within a unified infrastructure, enabling simultaneous acquisition and transmission of environmental data \cite{Beam}. Despite its advantages, this integration poses challenges, such as balancing the trade-off between communication and sensing, efficiently managing spectrum resources, and handling the increased system complexity \cite{kazemi2019k}.

Furthermore, ISAC utilizes the same RF spectrum for both communication and sensing, minimizing the need for dedicated frequency bands for each function. This approach requires strict frequency allocation controls to prevent interference and signal degradation \cite{RIS}. Dynamic allocation and spectral coordination techniques enable real-time adjustment of resource distribution based on demand. Additionally, energy consumption is optimized, as simultaneous communication and sensing necessitate a sophisticated processing infrastructure \cite{Levenberg}. 
While the integration of communication and sensing systems provides significant advancements in spectral efficiency and integrated functionality, it also introduces technical challenges that demand careful resource and performance management. As sensing technologies become more refined and sensitive, they consume a portion of the frequency spectrum and processing power, thereby reducing the bandwidth available for communication \cite{43}.

An important consideration is that ISAC system complexity increases with the inclusion of multifunctional infrastructures, such as RIS and adaptive beamforming algorithms, which require precise control and advanced processing to dynamically adjust RF signal characteristics according to environmental and operational conditions \cite{Beamformin2021}. Additionally, the synchronization between communication and sensing modules necessitates sophisticated control systems capable of real-time operation to effectively manage these interactions \cite{42}.

Building on these complexities, the need for 6G communication networks to provide innovative solutions for spectral resource management and interference mitigation underscores the importance of developing mathematical models and algorithms that integrate RIS and beamforming in massive MIMO systems with ISAC technology \cite{payal2013comparative, zayani2019efficient}. Accordingly, this study is motivated by the proposal of a novel ISAC system design that outperforms traditional systems by offering low computational complexity, instantaneous responses, centralized processing at the base station (BS), and high energy efficiency.

{A concise summary of the advancements in the techniques utilized reveals significant progress over time. The initial prototype, presented in \cite{P1}, employed only BS and trilateration techniques for user sensing, leveraging Zigbee communication protocols (IEEE 802.15.4). This approach demonstrated efficiency but required communication among four BSs to calculate user positioning. Subsequently, \cite{P2} introduced a more structured methodology for ISAC systems also incorporating trilateration techniques. In \cite{P3}, a notable advancement involved adapting the system from multiple BSs to a single BS with two RIS, significantly reducing energy consumption by eliminating the need for three additional BSs for sensing and communication. Further refinements were observed in \cite{P4}, where a system utilizing a single BS and one RIS integrated the time-division duplex (TDD) technique, enabling spectrum sharing between sensing and communication functionalities. Finally, \cite{P5} proposed an innovative approach aimed at optimizing the ISAC system, yielding promising results, albeit with the requirement of multiple frequencies spread required by space-time coding metasurface (STCM).} 

The paper's structure is organized as follows. Section \ref{sec:2} provides an overview of the general scenario, including channel definitions and the key technologies employed. Section \ref{sec:3} details the innovative implementation of the proposed ISAC system, highlighting the integration of sensing and communication functionalities, as well as a comparative analysis of computational complexity with existing approaches in the literature. In Section \ref{sec:4}, we present a probabilistic model for error occurrence and analyze the success rate of the proposed system. Section \ref{sec:5} discusses the simulation results, and finally, Section \ref{sec:6} concludes the paper.





\section{Preliminaries}\label{sec:2}

This section outlines the channel model used in the system and reviews the key technologies and methods essential for implementing the proposed ISAC system, such as MIMO techniques, adaptive beamforming, RIS technology, and a technical complexity analysis of ISAC systems.

\subsection{Channel Model}
\rf{The multipath fading effect is a common phenomenon in the propagation of electromagnetic waves in wireless channels, particularly in urban or indoor environments where multiple propagation paths coexist. This effect arises from the interaction of waves with environmental elements such as buildings, trees, and other structures, leading to reflections, diffractions, and scattering. Consequently, the received signal comprises multiple components with varying amplitudes, phases, and time delays, which can result in either constructive or destructive interference.}

\rf{Although solid obstacles might appear to be impenetrable barriers for electromagnetic waves, most materials do not entirely block wave propagation. Electromagnetic waves can pass through these objects via mechanisms such as partial transmission through dielectric materials and diffraction. For dielectric materials, a significant portion of the incident wave's energy is transmitted into the material, while only a small fraction is reflected. Diffraction, on the other hand, enables waves to bypass around obstacles, allowing propagation along paths that would otherwise be obstructed under LoS conditions.}

\rf{This behavior contrasts sharply with the interaction of electromagnetic waves with good conductor materials. According to the theory of the skin effect, the electromagnetic field of the wave cannot deeply penetrate a good conductor. Instead, the field remains confined to a very thin surface layer and is quickly reflected back into the environment. In the case of perfect conductors, wave penetration does not occur at all, and the electromagnetic wave is completely reflected.}

\rf{As expected, good conductor metallic objects are essential for dictating reflections within the environment, enabling the reflection necessary for sensing tasks. In contrast, other materials primarily contribute to multipath effects, which can be modeled by any fading channel, like the Rice channel. }

\subsubsection{Rice fading channel}
Consider a generic wireless communication channel represented by \textbf{G} $\in \mathbb{C}^{M_1 \times M_2}$, where $g_{m_1,m_2} \triangleq   \left[\textbf{\,G\,}\right]_{m_1,m_2}$ denotes the channel matrix element for the link between two arrays ($M_1$ and $M_2$) with $M_1$ and $M_2$ antennas, respectively.

Each element $g_{m_1,m_2}$ of the subchannel matrix $\mathbf{G}$ is described by a stochastic multipath fading model (${\underline{g}}_{m_1,m_2}$), scaled by a path loss coefficient ($\beta_{m_1,m_2}$). This is expressed as:
\begin{equation}
g_{m_1,m_2} = \beta_{m_1,m_2} \, {\underline{g}}_{m_1,m_2},
\end{equation}
where the path loss coefficient $\beta_{m_1,m_2}$ is given by:
\begin{equation}\label{pathloss}
\beta_{m_1,m_2} = {\frac{A \lambda^2}{(4\pi \alpha_{m_1,m_2})^2}},
\end{equation}
where $A$ represents the antenna gain, $\lambda$ is the wavelength of the propagated signal, and $\alpha_{m_1,m_2}$ denotes the distance between the $m_1$-th antenna of $M_1$ and the $m_2$-th antenna of $M_2$.

The multipath fading model is based on the Rice distribution, incorporating both LoS and NLoS components, represented as:
\begin{equation}
{\underline{g}}_{m_1,m_2} = \sqrt{\frac{\epsilon_{\text{rb}}}{\epsilon_{\text{rb}} + 1}} \bar{{g}}_{m_1,m_2} + \sqrt{\frac{1}{\epsilon_{\text{rb}} + 1}}\tilde{{g}}_{m_1,m_2},
\label{eq:7}
\end{equation}
where $\epsilon_{\textsc{rb}}$ is the Rice factor for the link $M_1$-$M_2$. The LoS component $\bar{{g}}_{m_1,m_2}$ represents the deterministic directional channel model, calculated as:
\begin{equation}\label{eq:svm}
\bar{{g}}_{m_1,m_2}\left(\mathbf{s}_{m_1},\mathbf{s}_{m_2}\right) = e^{-i\frac{2\pi}{\lambda} \left\| \mathbf{s}_{m_1} - \mathbf{s}_{m_2}\right\|},
\end{equation}
where $\left\|\mathbf{s}_{m_1} - \mathbf{s}_{m_2}\right\|$ represents the distance from the $m_2$-th antenna of $M_2$ to the $m_1$-th antenna of $M_1$.

The NLoS component introduces random dispersion due to various environmental factors. Each sample is modeled as:
\begin{equation}
\label{dist_Ray}
\tilde{{g}}_{m_1,m_2} = x_{m_1,m_2} + j \, y_{m_1,m_2},
\end{equation}
where $x_{m_1,m_2}$ and $y_{m_1,m_2}$ are the in-phase and quadrature components of the signal envelope, respectively. Both $x_{m_1,m_2}$ and $y_{m_1,m_2}$ are independent Gaussian random variables with zero mean and variance $\sigma^2$, such that $\sigma^2_X = \sigma^2_Y = \sigma^2$.

\subsection{Adaptive beamforming applied to the MIMO {system}}\label{sec:bmimo}

The advantages of MIMO communication systems are widely acknowledged by both industry and the telecommunications research community. These systems have become increasingly sophisticated in modern networks due to their ability to enhance spectral efficiency and increase data rates. By employing multiple antennas at both the transmitter and receiver, as well as supporting multiple users, MIMO enables the simultaneous transmission and reception of independent data streams. Additionally, MIMO technology optimizes communication resources and enhances robustness against interference and fading.

To further improve communication in MIMO systems, adaptive beamforming techniques are employed to maximize signal to noise ratio (SNR) and system capacity. Beamforming dynamically adjusts transmitted signals to align with channel conditions by utilizing precoders at the transmitter and combiners at the receiver. These techniques effectively minimize interference, mitigate noise, and improve power efficiency at the receiver.

Optimization through beamforming can be represented by an optimal weight matrix, $\mathbf{w}_{\text{opt}}\in \mathbb{C}^{M \times 1}$, which is determined by
\begin{equation}
    \mathbf{w}_{\text{opt}} = \arg\max_{\mathbf{w}} \left| \mathbf{h}_k^H \mathbf{w} \right|^2,
    \label{OtimBeamMIMO}
\end{equation}
in which $\mathbf{h}_k \in \mathbb{C}^{M \times 1}$ represents the $k$-th column vector of the channel matrix $\mathbf{H} \in \mathbb{C}^{M \times K}$, where $M$ is the number of base station (BS) antennas, and $K$ is the number of users. Specifically, $\mathbf{H} = [\mathbf{h}_1, \, \mathbf{h}_2, \, \mathbf{h}_3, \, \ldots, \, \mathbf{h}_K]$ represents the direct channel component between the BS and the users. The symbol $^H$ denotes the Hermitian transpose (conjugate transpose) of a matrix. 

Classical Maximum Ratio Transmission (MRT) beamforming techniques are employed, focusing on their mathematical structure and use in advanced communication systems. 
The MRT is a precoding technique applied at the transmitter to maximize the SNR at the receiver. It achieves this by adjusting the weight vector $\mathbf{w} \in \mathbb{C}^{M \times 1}$ to align the transmitted signals in phase. This method is particularly well-suited for single-user applications. The mathematical expression for the beamforming vector in MRT is given by

\begin{equation}
    \mathbf{w}_{\text{MRT}} = \frac{\mathbf{h}^H}{\|\mathbf{h}\|}.
    \label{eq:MRT}
\end{equation}

The application of MRT is recommended in scenarios with a single-user LoS connection between the transmitter and receiver, as MRT always ensures optimal performance in terms of SNR, since it operates under such an AWGN channel.

On the other hand, the MRC combiner is applied at the receiver side. It combines the signals received by multiple antennas to maximize SNR. The weight vector is adjusted to align the phases of the received signals, ensuring that all components contribute constructively to the final signal's intensity. In essence, the key distinction between MRC and MRT lies in their respective side 
of application: MRT is implemented at the transmitter, while MRC is applied at the receiver.

Thus, the performance analysis of the MIMO system can be conducted using metrics such as SNR for single-user systems, SINR for multi-user systems, and channel capacity. The optimized SNR is given by Eq. \eqref{DesempMIMO}.

\begin{equation}
    \text{SNR}_{\text{opt}} = \frac{P_{\text{trans}} \cdot \left| \mathbf{h}_k^H \mathbf{w}_{\text{opt}} \right|^2}{\sigma^2_n},
    \label{DesempMIMO}
\end{equation}
where $P_{\text{trans}}$ is the transmission power and the noise power is measured by,
 \begin{equation}
    \sigma^2_n = 10^{\frac{N_0}{10} - 3} \textnormal{B}{,}
    \label{Pruido}
\end{equation}
here, $ N_0 $ represents the noise power spectral density, typically set at $-204$ dBm/Hz, which corresponds to the thermal noise level under ideal room temperature conditions of approximately $ 17^\circ \text{C} $. Meanwhile, $ B $ denotes the system bandwidth.

For multi-user systems, the SINR is formulated as:

\begin{equation}
\textnormal{SINR}_k = \frac{\left|  \mathbf{h}_{k}^H \mathbf{w}_k \right|^2}{\sum_{i=1, i \neq k}^{K} \left| \mathbf{h}_{i}^H\mathbf{w}_k \right|^2 + {\sigma_n^2}}{.}
\label{eq:sinr}
\end{equation}

The integration of RIS into MIMO systems significantly enhances system capacity and signal quality, even in challenging scenarios with multipath propagation and obstacles. These systems are discussed in detail in Section \ref{sec:ris}.

\subsection{RIS}\label{sec:ris}

RIS is emerging as an innovative technology in wireless communications, designed to transform signal propagation environments into controlled and adaptable scenarios. It consists of two-dimensional arrays of passive electronic elements capable of dynamically manipulating the properties of incident electromagnetic waves, such as phase, amplitude, and polarization, through {optimized} control of each surface element. These surfaces can reflect, refract, or absorb electromagnetic signals in a programmable manner, optimizing signal propagation in complex and heterogeneous environments.  

In MIMO systems, RIS is utilized to shape the propagation channel optimally, resulting in significant enhancements in transmission capacity, SNR, and network coverage.  

The mathematical modeling of RIS involves representing its surface elements as complex reflection coefficients, which are dynamically adjusted to control the interaction between the surface and the incident waves. Each RIS element is modeled as a programmable reflector, with properties tailored to modify the direction, phase, of the reflected signal. The reflection matrix $\mathbf{\Theta} \in \mathbb{C}^{N \times N}$, for a RIS composed of $ N $ elements, is defined as $\mathbf{\Theta} = \text{diag}(\boldsymbol{\theta} ),$ 
in which, $\boldsymbol{\theta} = [\theta_1\,\,\,\theta_2\,\,\,\dots\,\,\,\theta_N] \in \mathbb{C}^{1 \times N}$, where  $\theta_i = \beta_i e^{j\phi_i}$  represents the reflection coefficient of the $ i $-th element. In this expression, $ \beta_i \in [0, 1] $ is the amplitude modulation coefficient, responsible for adjusting the intensity of the reflected signal, and $ \phi_i \in [0, 2\pi) $ is the adjustable phase, which determines the phase delay imposed on the reflected signal.

 Furthermore, the RIS channel model can be described by combining multiple paths, including both direct and reflected channels. Similar to the MIMO system, the channel matrix retains the same dimensions as the direct channel presented in Section \ref{sec:bmimo}. Specifically, $ \mathbf{H} = [\mathbf{h}_1, \, \mathbf{h}_2, \, \mathbf{h}_3, \, \ldots, \, \mathbf{h}_K] \in \mathbb{C}^{M \times K} $. By incorporating the contributions of RISs, the model is structured as represented in Eq. \eqref{ModCanRIS}.

\begin{equation}
    \mathbf{h}_k = \mathbf{h}_{\text{LoS}_k} + \sum_{k=1}^{K} \mathbf{G}_{\text{BS-RIS}_i}^H \mathbf{\Theta}_i \mathbf{f}_{\text{RIS}_i-k},
    \label{ModCanRIS}
\end{equation}
in which, $\mathbf{H}_{\text{LoS}} = [\mathbf{h}_{\text{LoS}_1}, \, \mathbf{h}_{\text{LoS}_2}, \, \mathbf{h}_{\text{LoS}_3}, \, \ldots, \, \mathbf{h}_{\text{LoS}_K}] \in \mathbb{C}^{M \times K}$ represents {a pure LoS} direct channel between the BS and the $K$ users. The matrix $\mathbf{G}_{\text{BS-RIS}_i} \in \mathbb{C}^{N_i \times M}$ denotes the channel between the BS and the $i$-th RIS, while $\mathbf{f}_{\text{RIS}_{i}\text{-}k} \in \mathbb{C}^{N_i \times 1}$ represents the channel between the $i$-th RIS and user $ k $. Additionally, $ \mathbf{\Theta}_i \in \mathbb{C}^{N_i \times N_i} $ is the diagonal reflection matrix of the $ i $-th RIS.

For LoS paths, where the signal propagates directly from the BS to the points of interest without significant obstacles, the SNR is typically higher due to reduced signal attenuation. {This characteristic is particularly advantageous because propagation loss is minimized, and is typical of open environments or scenarios with direct visibility.}

However, in NLoS paths, where the direct path is heavily obstructed, it becomes necessary to implement RIS to strengthen the wireless communication link. In such cases, the signal is reflected by the RIS before reaching the points of interest, and the SNR is influenced by the channel gains of each segment along the path, including the interaction with the smart surfaces.

The formulations for SNR and SINR metrics in NLoS scenarios are adapted from Eqs. \eqref{DesempMIMO} and \eqref{eq:sinr}, resulting in the following expressions, respectively:
\begin{align}
    \text{SNR}_{\text{NLoS}} &=\frac{\left| \left(\mathbf{\mathbf{G}}^H \mathbf{\Theta}_i \mathbf{h}_{k} \right)^H \mathbf{w}_k \right|^2}{ {\sigma_n^2}};
    \label{SNRNLoS}\\
\textnormal{SINR}_{\text{NLoS}} &= \frac{\left| \left( \mathbf{\mathbf{G}}^H \mathbf{\Theta}_i \mathbf{h}_{k} \right)^H \mathbf{w}_k \right|^2}{\sum_{j=1, j \neq k}^{K} \left| \left( \mathbf{G}^H \mathbf{\Theta}_i \mathbf{h}_{j} \right)^H \mathbf{w}_k \right|^2 + {\sigma_n^2}}.\label{SINRNLoS}
\end{align}

This channel gain is influenced by factors such as the path loss associated with the distance between the BS and the RIS, the distance between the RIS and the points of interest, and the reflection coefficients of the RIS. Additionally, the application of adaptive beamforming enhances the system's SNR. The optimal weight vector,  $\mathbf{w}_{\text{opt}}$, enables dynamic steering of the signal beams by adjusting the phase and amplitude of the signals emitted by the BS antennas to align with the composite channel, as modeled by Eq. \eqref{ModCanRIS}. This alignment maximizes signal power in the desired directions and minimizes interference in undesired ones, resulting in improved SNR or SINR at the points of interest, even in scenarios with a high prevalence of NLoS paths.

\subsubsection{\textit{Beanforming} passivo}

Passive beamforming is an innovative and promising technique for future 6G networks that intelligently reconfigures the reflective elements of the RIS to optimize signal propagation in wireless communication systems. Unlike active beamforming, where signals are processed and amplified at the transmitters and receivers, passive beamforming manipulates the reflections of incident signals through programmable reflective elements. Each RIS element can dynamically adjust the phase of the reflected signals, enabling controlled and directed signal propagation. This approach is particularly attractive due to its energy efficiency and low cost, as it eliminates the need for power amplifiers or active signal processing.

This paper adopts a passive beamforming approach, as documented in the literature, which demonstrates excellent performance in single-user scenarios. By integrating optimal precoders and combiners with passive beamforming for multi-user systems, significant simultaneous gains are achieved for all users. Consequently, the application of passive beamforming in massive MIMO systems aided by RIS emerges as a viable, suboptimal solution for multi-user scenarios.

Based on detailed channel knowledge, Eq. \eqref{theta_opt} defines the adjustments required to align the phases of the signal components, in order to maximize the power of the resulting signal in the direction of the intended user
\begin{equation}\label{theta_opt}
\boldsymbol{\theta}_k=\exp{[-j\, \arg(\mathbf{w}_k^H \mathbf{G}^H \text{Diag}(\mathbf{h}_k))]},
\end{equation}
here, $\boldsymbol{\theta}_k \in \mathbb{C}^{1 \times N}$ represents the passive beamforming configuration of the $N$ reflecting elements of the RIS for user $k$, where the active ($\mathbf{w}_k$) and passive beamforming vectors are interdependent. To obtain a suboptimal configuration of the RIS elements, the passive beamforming vector can be initialized randomly, the corresponding active beamforming vector computed, and then the passive vector updated accordingly. This process can be iterated until convergence is achieved.

\section{Three-dimensional Localization Modelling}\label{sec:3}

The three-dimensional localization system {applied} in this work is an innovative technique that integrates a BS and two RISs into an ISAC system. This is designed to accurately determine the location of two specific points within a three-dimensional environment using trilateration sensing techniques combined with time multiplexing and data communications. The proposed framework is described in detail in the following subsections.

\subsection{Sensing system model}

The three-dimensional environment is represented by a Cartesian coordinate system $(x, y, z)$, where the position of the BS is denoted as $\mathbf{P}_{BS} = (BS_x, BS_y, BS_z)$. The BS emits the primary signal for communication and environmental sensing purposes. The positions of the two RISs are fixed at $\mathbf{P}_{RIS_1} = (R1_x, R1_y, R1_z)$ and $\mathbf{P}_{RIS_2} = (R2_x, R2_y, R2_z)$, and these RISs reflect the sensing signal back to the BS. The signal originates from the BS, reflects off the user in all directions, and is subsequently redirected to the BS by the two RISs. The user location coordinates (User Equipment - UE) are represented by $\mathbf{P}_{\textnormal{UE}} = (\textnormal{UE}_x, \textnormal{UE}_y, \textnormal{UE}_z)$.



\begin{figure}
    \centering
\includegraphics[width=0.75\linewidth]{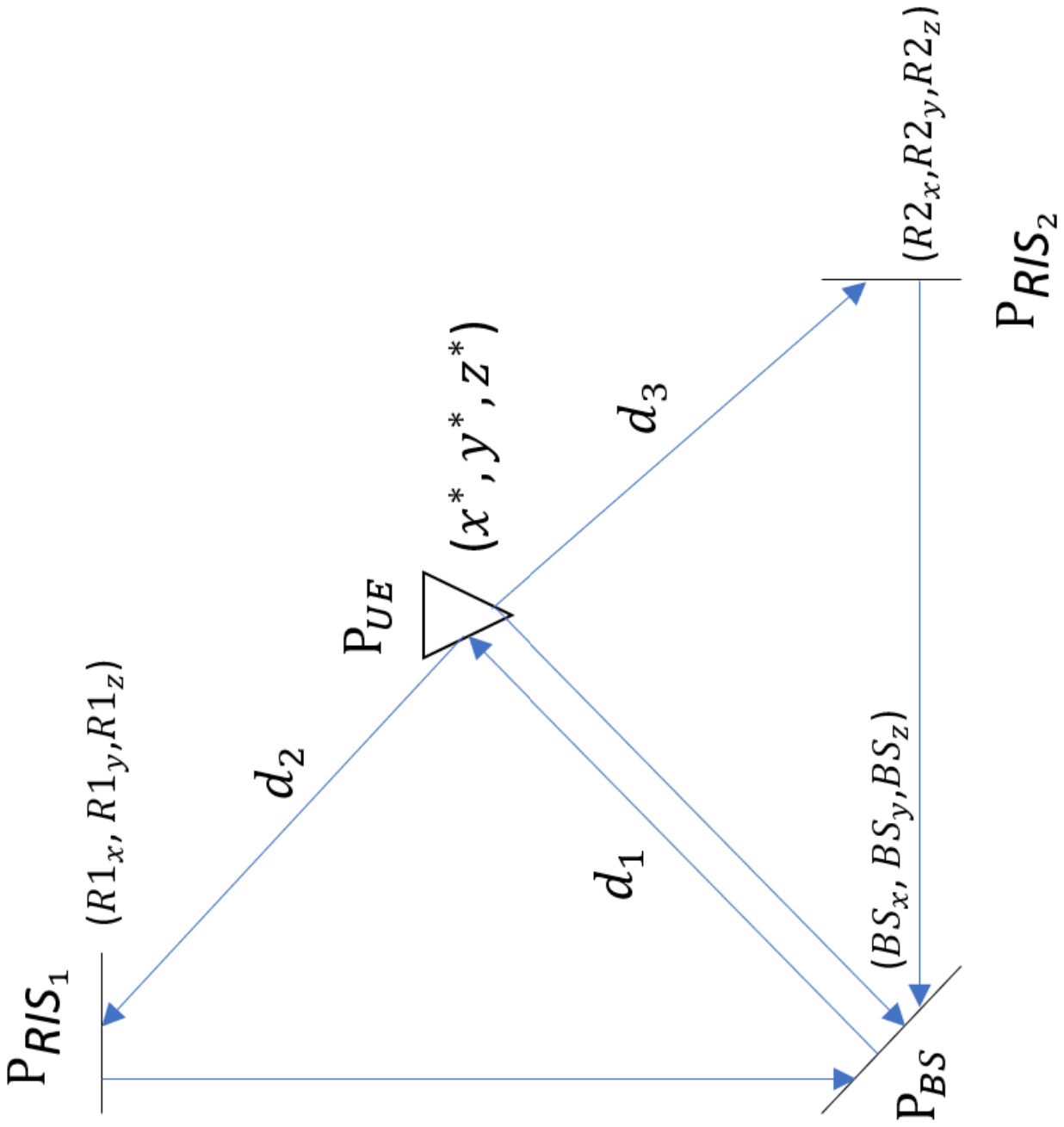}
    \caption{ISAC system model composed by one BS and 2 RIS.}
    \label{sismod}
\end{figure}

The user localization process is performed by the BS through exclusive sensing signals {(Sensing Frame, Figure \ref{fig:tdd})}. These signals are transmitted to cover a region {with a wide beam }and, upon interacting with environmental elements, are reflected omnidirectionally before returning to the BS through three distinct paths. These paths are described below and illustrated in Figure \ref{sismod}:

\textbf{Path 1:} The signal is reflected directly by the user and returns via the direct path, where $ d_1 $ represents the distance from the user to the BS;  

\textbf{Path 2:} The signal is reflected towards RIS$_1$, which subsequently reflects it towards the BS, where $ d_2 $ denotes the distance from the user to RIS$_1$; 

\textbf{Path 3:} The signal is reflected towards RIS$_2$, which also reflects it towards the BS, where $ d_3 $ represents the distance from the user to RIS$_2$.   

Reflections enable the BS to determine the location of the user information using the trilateration technique. However, employing three anchors results in two possible location points for the user. {Our system is designed to provide the user's height (inside the Permission and Response Frame, Figure \ref{fig:tdd}) as additional data to BS identify which of these two candidate points corresponds to the actual user location}.  

{The channel model for sensing process can initially be represented by the first reflected path, which is characterized by the channel $\mathbf{h}_{\text{BU}} \in \mathbb{C}^{M \times 1}$,modeled as:}
\begin{equation}
    \mathbf{h}_{\text{BU}} =  \mathbf{h}_{\text{BS}}^{H}\, \mathbf{\Theta}_{\text{UE}}\,\mathbf{h}_{\text{UR}},
    \label{CanalLoS1}
\end{equation}
{in which  $\mathbf{h}_{\text{BS}}$ $\in \mathbb{C}^{1 \times M}$ represents the channel from the BS active beamforming precoding with sufficient aperture to encompass the user positioned at \(\mathbf{P}_{\textnormal{UE}}\). \(\mathbf{\Theta}_{\text{UE}} \in \mathbb{C}^{1 \times 1}\) models the random reflection and absorption of the sensing signal by the user, while \(\mathbf{h}_{\text{UR}} \in \mathbb{C}^{1 \times 1}\) corresponds to the channel from the user, where the signal is reflected omnidirectionally.  

The other two paths are modeled by \(\mathbf{h}_{\text{URB}_i} \in \mathbb{C}^{1 \times M}\), where \(i = \{1, 2\}\), representing the signal reflection by the user towards the \(i\)-th RIS, which then redirects the signal to the BS using passive beamforming, with Eq. \eqref{theta_opt}. The channel model for this reflection is characterized by:
 \begin{equation}
    \mathbf{h}_{\text{URB}_i} =  \mathbf{h}_{\text{UR}_i}^{H}\, \mathbf{\Theta}_{\text{RIS}_i}\,\mathbf{H}_{\text{RB}_i},
    \label{CanalNLoS}
\end{equation}
\rf{where, \(\mathbf{h}_{\text{UR}_i} \in \mathbb{C}^{N_i \times 1}\) represents the channel between the user and RIS\(_i\);   \(\mathbf{H}_{\text{RB}_i} \in \mathbb{C}^{N_i \times M}\) describes the channel between RIS\(_i\) and the BS; and $\mathbf{\Theta}_{\text{RIS}_i}=\text{Diag}([e^{j {\theta}_{\text{RIS}(i,1)}},\,e^{j {\theta}_{\text{RIS}(i,2)}},...,\,e^{j {\theta}_{\text{RIS}(i,j)}},...,\,e^{j {\theta}_{\text{RIS}(i,N_i)}}])$, ${\theta}_{\text{RIS}(i,j)} \in[0,2 \pi)$. $\mathbf{\Theta}_{\text{RIS}_i} \in \mathbb{C}^{N_i \times N_i}$ denotes the pre-configured reflection model of RIS\(_i\) for directional reflection of the sensing signal.}

{The system architecture is designed to achieve a broad coverage area exclusively for the sensing signal. According to MIMO system theory, increasing the number of antennas enhances array gain. However, this comes at the expense of reduced coverage. To balance coverage and array gain, the system is designed with three base station (BS) antennas ($M=3$) and three elements per RIS ($N_i=3$).}


Additionally, the communication and sensing are structured as a system similar to TDD in which specific time slots are allocated to each function as illustrated in Figure~\ref{fig:tdd}.

\begin{figure}
    \centering
    \includegraphics[width=0.95\linewidth]{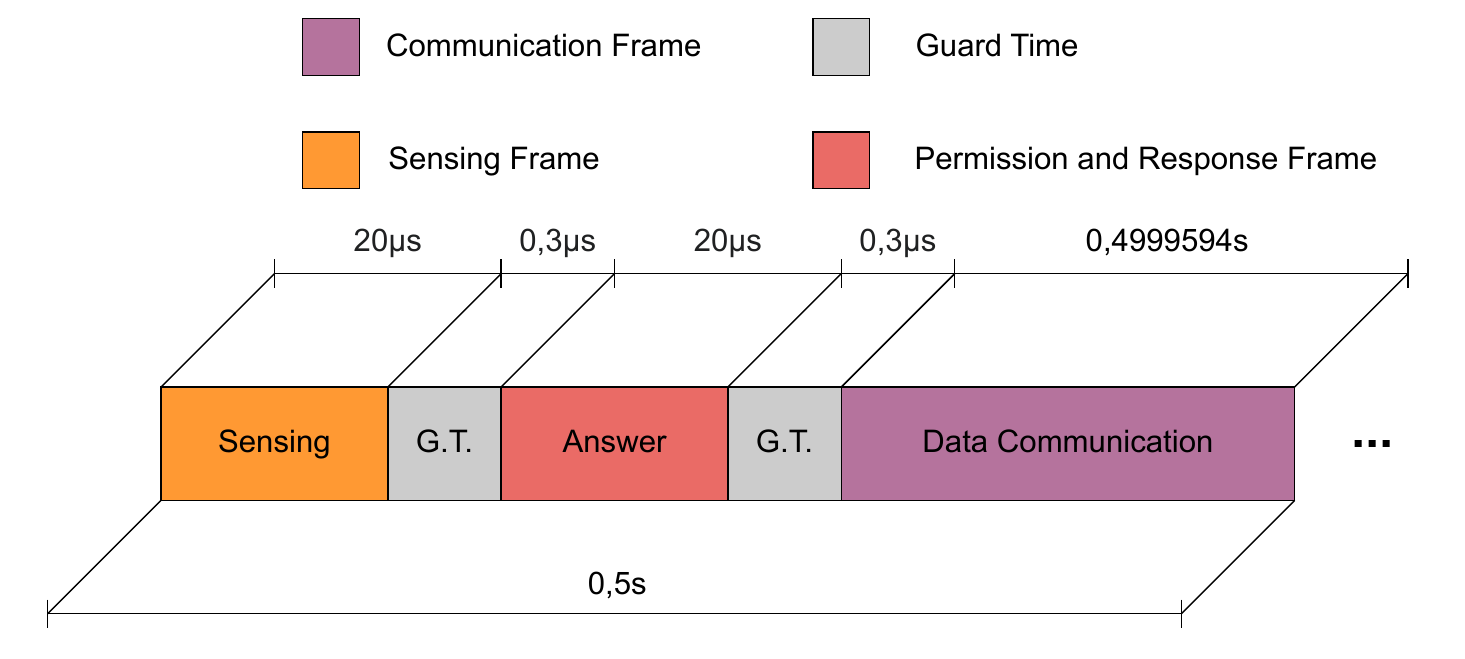}
    \caption{Proposed ISAC Frame Structure with Time Allocation.}
    \label{fig:tdd}
\end{figure}

\rf{In this approach, communication is allocated the majority of the time, while sensing is assigned a small fraction. The sensing process is designed to periodically monitor a pre-defined user and operates with low data rate requirements. This allocation strategy is intentionally designed to avoid conflicts between communication and sensing signals.}

\rf{In addition, the system parameters were analyzed to ensure communication and sensing performance. Using a modulation scheme {of order} $16$, which corresponds to $4$ bits per symbol, with $500$ KHz bandwidth providing a bit rate of $2$ Mbits/s. For the transmission of $40$ bits, the total transmission time was calculated as $20$ microseconds. Additionally, the round-trip time for a signal to travel a distance of $100$ meters was determined to be $0.3$ microseconds. These parameters highlight the system's operation timing,  as a cycle period of $0.5$ seconds leaves $0.4999494$ seconds available for data communication.}

{To ensure that the sensing signal data is securely addressed to a specific user, we propose the following frame structure for the sensing signal.}
\begin{table}[H]
\centering
\caption{ Sensing Frame Structure  - 40 bits.}
\begin{tabular}{|p{2.5cm}|c|>{\raggedright\arraybackslash}p{4.5cm}|}
\hline
\textbf{Term} & \textbf{Bits} & \textbf{Description} \\
\hline
Sensing preamble & 5 & Standard sequential bit structure for sensing, recognized by all devices in the system. \\
\hline
User ID  & 20 & Unique identifier assigned to each user. \\
\hline
Timestamp & 7 & Indicates the measurement instant; can also serve as a cyclic counter for temporal ordering. \\
\hline
Cyclic Redundancy Check (CRC) & 8 & Used to verify the integrity of transmitted data.. \\
\hline
\end{tabular}
\end{table}

{The Permission and Response Frame is dedicated solely to the user's response, identified by the User ID within the Sensing Frame. If the UE agrees to be monitored, it sends a positive response frame back to the BS, including both its permission and height information. Once the BS acknowledges the UE’s response, it is ready to initiate directive communication.}

\subsection{\ta{Sensing Operation Mode}}
The proposed configuration for this new mode of operation significantly reduces the sensing power requirements in an ISAC problem, as it eliminates the need for three additional BSs compared to traditional trilateration sensing \cite{P2}. While the power of the RISs is {considered in} the system, their operating power is negligible compared to that of the BS. Consequently, the total power required for the efficient operation of the system, considering both the BS and the RIS, is expressed as the sum of the involved powers, as detailed in Eq. \eqref{PotSist1},
\begin{equation}
    P_{\text{total}} = P_{\text{BS}_{sens.}} + \sum_{i=1}^{2} P_{\text{RIS}_i},
    \label{PotSist1}
\end{equation}
 here \( P_{\text{RIS}_i} \) represents the power consumed by the \( i \)-th RIS to reflect and adjust the signal beam, and \( P_{\text{BS}_{sens.}} \) denotes the sensing power of the BS. The latter is calculated by considering the power radiated by the BS towards the points of interest, taking into account the channel characteristics, as well as the attenuation and reflection properties. Its value must be sufficient to ensure that the sensing signal returns to the BS with acknowledgement. The expression for \( P_{\text{BS}_{sens.}} \) is given by Eq. \eqref{Ptrans},
\begin{equation}
    P_{\text{BS}_{sens.}} = P_{\text{BS}} \cdot \left|  \mathbf{h}_{\text{BU}}  \mathbf{w}_{1Q} \right|^2
    \label{Ptrans},
\end{equation}
in which \( P_{\text{BS}} \) represents the initial transmit power of the BS, and \( \left| \mathbf{h}_{\text{BU}} \mathbf{w}_{1Q} \right|^2 \) denotes the BS channel gain configured by the precoder (\( \mathbf{w} \)) to provide coverage specifically for a wide region in the first quadrant of the Cartesian space.

\subsection{Sensing equations}

In this section, the localization problem is formalized algebraically. The objective is to develop a mathematical model for the precise determination of the coordinates of the UE in space. To achieve this, the equations of spheres are utilized, as they serve as a fundamental tool for representing the spatial boundaries within which objects can be located.  

First, the equations of spheres centered at the points $(BS_x, BS_y, BS_z)$, $(R1_x, R1_y, R1_z)$ and $(R2_x, R2_y, R2_z)$ are presented, as pointed in Figure \ref{sismod}. These equations are organized into the system below,
\begin{align}
&(x - BS_x)^2 + (y - BS_y)^2 + (z - BS_z)^2 = d_1^2; \label{eq1} \\
&(x - R1_x)^2 + (y - R1_y)^2 + (z - R1_z)^2 = d_2^2; \label{eq2}\\
&(x - R2_x)^2 + (y - R2_y)^2 + (z - R2_z)^2 = d_3^2, \label{eq3}
\end{align}
in which $\mathbf{d}=[d_1,d_2,d_3]^T$ represents the distances from the UE's point of interest to the anchors.

The combination of these spherical equations forms a system of equations that, when solved, determines the exact location of the UE. This method, commonly referred as trilateration, is extensively used in geolocation applications and navigation systems, delivering accurate and reliable results.  

Solving this system yields specific values for the variables $x^*$, $y^*$, and $z^*$, which correspond to the precise position of the user in three-dimensional space. The detailed development of the exact analytical expressions for determining the user's position within the proposed system model is presented in {the following}.

\subsubsection{Optimization Problem}\label{sec:nl}

The localization problem described above can be formulated as a nonlinear optimization problem, with the objective of minimizing the squared error between the calculated distances and the measured distances from the points of interest to the reference points, which include the BS and the RIS.  

Specifically, the localization problem is expressed as an optimization problem aimed at minimizing the total error $ E(x, y, z)$, which represents the sum of the squared differences between the actual distances and measured distances for each sphere defined. The objective function to be minimized is provided in Eq. \eqref{MSENlinear},
\begin{align}
    \min_{x, y, z} E(x, y, z) = \min_{x, y, z} \left[ \right.&\sum_{i=1}^{3} \left( \right.d_i^2 - \left[\right.(x - P_{i_x})^2 + (y - P_{i_y})^2 \notag \\
    &  + (z - P_{i_z})^2\left. \right] \left.\right)^2 \left.\right],\label{MSENlinear}
\end{align}
here, \( d_i \) represents the actual distances from the points of interest to the anchors positions, \((P_{i_x}, P_{i_y}, P_{i_z})\) with $i=1,2,3$, denotes the central coordinates of the three spheres defined by the locations of the BS and the two RIS, given respectively by \((BS_x, BS_y, BS_z)\), \((R1_x, R1_y, R1_z)\), and \((R2_x, R2_y, R2_z)\). In which the decision variable vector is defined as $\mathbf{p} = [x, y, z]^T$.

{The total error evaluate localization accuracy by comparing the the actual distances $d_i$ to the estimated distances vector 
$\mathbf{c}(\mathbf{p})\!=\![{c}_1(\mathbf{p}),\! {c}_2(\mathbf{p}),\! {c}_3(\mathbf{p})]^T$ in which ${{c}_i(\mathbf{p})}=(x - P_{i_x})^2 + (y - P_{i_y})^2+ (z - P_{i_z})^2$ corresponds to the square of the Euclidean distance between the point of interest \(\mathbf{p}\) and the positions references, as}
\begin{equation}
    \hspace{-2mm}\begin{bmatrix} 
   c_1(\mathbf{p}) \\ 
    c_2(\mathbf{p}) \\ 
    c_3(\mathbf{p})
    \end{bmatrix}\!\!=\!\! \begin{bmatrix} 
   \! (x - BS_x)^2 + (y - BS_y)^2 + (z - BS_z)^2\! \\ 
   \! (x - R1_x)^2 + (y - R1_y)^2 + (z - R1_z)^2\! \\ 
   \! (x - R2_x)^2 + (y - R2_y)^2 + (z - R2_z)^2 \!
    \end{bmatrix}\!\!.
    \label{MatCustoDef}
\end{equation}

The goal of the optimization problem is to determine the coordinates \(\mathbf{p}\) that minimize the total error, defined as the sum of the squared differences between the calculated and observed distances. This is represented by the total error equation provided in Eq. \eqref{ErroTotCust},
\begin{equation}
    E(\mathbf{p}) = \sum_{i=1}^{3} \left( d_i^2 - {c}_i(\mathbf{p}) \right)^2.
    \label{ErroTotCust}
\end{equation}
this value encapsulates the discrepancy between the measured distances and the distances estimated by the model, forming the foundation for calculating the total error to be minimized. {Thus, the nonlinear optimization problem can be expressed in matrix form as shown in Eq. \eqref{Simplf}},


\begin{align}\label{Simplf}
    \min_{\mathbf{p}} \,\,\,\,\,& \left\| \mathbf{d} - \mathbf{c}(\mathbf{p}) \right\|^2\\
    s.t. \,\,\,\,\,&\mathbf{p}\geq0.\notag
\end{align}

To solve this nonlinear optimization problem, several numerical methods can be employed, such as the Nonlinear Least Squares (NLS) method, which aims to find a solution \(\mathbf{p}^*\) that minimizes the sum of squared errors. This approach uses nonlinear adjustment techniques, such as Gradient Descent, which iteratively updates the position vector \(\mathbf{p}\) until the gradient of the objective function \(E(\mathbf{p})\) approaches zero, indicating a minimum point.  
Another effective method is the Levenberg-Marquardt algorithm, a hybrid technique that combines the advantages of Gradient Descent and the Gauss-Newton method. It is particularly suitable for nonlinear adjustment problems, as it dynamically adjusts parameters to improve convergence \cite{payal2013comparative}.  
However, for this problem, we will solve the optimization problem analytically to obtain a solution with a \rf{constant complexity, \(\mathcal{O}(1)\)}, ensuring computational efficiency in user localization.

The solution to the problem can be achieved using a linearization method, as demonstrated in Appendix \ref{appendix:proof}. This approach transforms the original problem into an equivalent linear problem, which is given by,
\begin{align}\label{eqs_lineares}
    a_1 \, x  + b_1  \,y + c_1 \,z = D_1\notag\\
    a_2 \, x  + b_2  \,y + c_2 \,z= D_2\\
    a_3 \, x  + b_3  \,y + c_3 \,z= D_3,\notag
\end{align}
in which 
\begin{align}
    a_1=\textnormal{R1}_x - \textnormal{BS}_x\quad\quad
    b_1=\textnormal{R1}_y - \textnormal{BS}_y\quad\quad
    c_1=\textnormal{R1}_z - \textnormal{BS}_z\notag\\
    a_2=\textnormal{R2}_x - \textnormal{BS}_x\quad\quad
    b_2=\textnormal{R2}_y - \textnormal{BS}_y\quad\quad
    c_2=\textnormal{R2}_z - \textnormal{BS}_z\notag\\
    a_3=\textnormal{R2}_x - \textnormal{R1}_x\quad\quad
    b_3=\textnormal{R2}_y - \textnormal{R1}_y\quad\quad
    c_3=\textnormal{R2}_z - \textnormal{R1}_z,
\end{align}
\begin{align}
    D_1 = \frac{d_1^2 - d_2^2 + \textnormal{R1}_x^2 - \textnormal{BS}_x^2 + \textnormal{R1}_y^2 - \textnormal{BS}_y^2 + \textnormal{R1}_z^2 - \textnormal{BS}_z^2}{2}\notag\\
    D_2 = \frac{d_1^2 - d_3^2 + \textnormal{R2}_x^2 - \textnormal{BS}_x^2 + \textnormal{R2}_y^2 - \textnormal{BS}_y^2 + \textnormal{R2}_z^2 - \textnormal{BS}_z^2}{2}\notag\\
    D_3 = \frac{d_2^2 - d_3^2 + \textnormal{R2}_x^2 - \textnormal{R1}_x^2 + \textnormal{R2}_y^2 - \textnormal{R1}_y^2 + \textnormal{R2}_z^2 - \textnormal{R1}_z^2}{2}.
\end{align}

The solution to the linear problem presented in Eq. \eqref{eqs_lineares} is obtained using Cramer's Rule, resulting in:

\begin{align}
  x^* &= \frac{D_1 b_2 - D_2 b_1}{a_1 b_2 - a_2 b_1},\label{eq:placeholder_label1} \\
  y^* &= \frac{a_1 D_2 - a_2 D_1}{a_1 b_2 - a_2 b_1},\label{eq:placeholder_label2} \\
  z^* &= \textnormal{BS}_z \pm \sqrt{d1^2 - \left(x^* - \textnormal{BS}_x \right)^2 - \left(y^* - \textnormal{BS}_y \right)^2}.
\label{eq:placeholder_label3}
\end{align}

\subsection{ToA and AoA techniques}

The following methods are employed to determine the propagation time of the sensing signal within the system and to identify its corresponding propagation paths \rf{to properly proceed with the trilateration method.}

The ToA metric is calculated by the system to measure the propagation time of the sensing signal across the three paths depicted in Figure \ref{sismod}. {Although delays are typically derived from known distances, in practical scenarios the distances are unknown and must be estimated from the observed delays. This estimation can be performed, for example, by identifying the correlation peak between the transmitted and reflected signals. It is important to note that the estimated time delays include inherent errors, as they represent the true delays plus a certain estimation error $\epsilon$. Therefore, for the purposes of this work, we adopt the approximation that these estimated delays are sufficiently accurate for subsequent processing.}

{To model this estimation uncertainty more rigorously, the ToA is expressed as the sum of the true delay and an error term, $\epsilon_k$, with $k = 1, 2, 3$, accounting for random adversities encountered during signal propagation through the wireless channel}, including those induced by multipath effects. This error term is modeled as a Gaussian random variable with zero mean and variance {proportional to the channel gain, i.e., $\mathcal{N} \left( 0, \frac{|h|^2}{\text{SNR}} \right)$}, {in which $h$ denotes the channel coefficient. This modeling approach ensures a constant instantaneous SNR, independent of the specific realization of $h$, at the cost of introducing a statistical dependence between the noise and the channel.}

The transmission time for path $1$ (\(t_1\)) corresponds to the duration in which the signal is transmitted by the BS, reflected by the user, and returns directly to the BS. This round-trip time can be expressed as:
\begin{equation}
    t_1=2 \frac{d_1}{c}+\epsilon_1,
\end{equation}
here $c$ is the speed of light in a vacuum.

The transmission time for path 2 (\(t_2\)) corresponds to the duration in which the signal is transmitted by the BS, reflected by the user, and then propagated towards RIS\(_1\), which reconfigures the incident waves to return them to the BS. The total round-trip time for the signal is expressed as:
\begin{equation}
    t_2=\frac{d_1+d_2+d_{R1-BS}}{c}+\epsilon_2,
\end{equation}
in which $d_{R1-BS}$ corresponds to the distance between RIS$_1$ and the BS.

Similarly, the transmission time for path 3 (\(t_3\)) corresponds to the duration in which the signal is transmitted by the BS, reflected by the user, and then propagated towards RIS\(_2\), which reconfigures the incident waves to return them to the BS. The total round-trip time for the signal is expressed as:
\begin{equation}
    t_3=\frac{d_1+d_2+d_{R2-BS}}{c}+\epsilon_3,
\end{equation}
where $d_{R2-BS}$ represents the distance between RIS$_2$ and BS.

In this system model AoA is utilized to identify the origin of the propagation paths. \rf{It is essential to point out that complex AoA models are unnecessary in this context, as the primary requirement is to identify the origin of the propagation path (i.e., the RIS anchor from which the signal was reflected) to enable accurate trilateration. Such identification is fundamental to eliminate ambiguities during the trilateration process, with the receiving antenna at the BS inherently functioning as a path identification sensor}. The BS predicts the AoA through a simple and efficient technique that leverages the properties of the incoming signal.  

The BS is equipped with an antenna array (BS\(_1\), BS\(_2\), BS\(_3\), ... , BS\(_M\)), {using this structure, we implement spatial filtering. Since the positions of the RISs are fixed and known, the sensing signal received at the $M$ antennas of the BS is correlated with a steering vector oriented toward the direction of each RIS$_i$, allowing us to identify the $i$-th path. As a result, we can isolate the signals arriving from each RIS. By estimating the two reflected signals, the one with the lowest correlation to the steering vectors is attributed to the direct path from the UE. The procedure for identifying the signal paths is detailed in the following.}


\vspace{2mm}
\noindent\textbf{Path identification 2:} {The path associated with this signal corresponds to the one exhibiting the highest correlation with the steering vector from the BS toward RIS$_1$. In other words, the path is identified as the signal that maximizes the following spatial correlation expression:
\begin{equation}\label{sv2}
    \textnormal{Path2}=\max_{i\in{1,2,3}} \left| \mathbf{a}^H(\theta_{\text{ RIS1}}) \hat{x}_i \right|, 
\end{equation}
in which $\hat{x}_i$, with $i\in 1,2,3$ represents the three received signals from the three different paths, $\mathbf{a}^H(\theta_{\text{RIS1}})$ represents the steering vector corresponding to the expected direction of the direct path between RIS$_1$ and the BS. Thus, if $\hat{x}_j$ is the term that maximizes the Eq. \eqref{sv2}, it is classified as belonging to Path2.}

\vspace{2mm}
\noindent\textbf{Path identification 3:} {The path associated with this signal corresponds to the one exhibiting the highest correlation with the steering vector from the BS toward RIS$_2$. In other words, the path is identified as the signal that maximizes the following spatial correlation expression:
\begin{equation}\label{sv3}
    \text{Path3}=\max_{i\in{1,2,3}} \left| \mathbf{a}^H(\theta_{\text{ RIS2}}) \hat{x}_i \right|, 
\end{equation}
in which $\mathbf{a}^H(\theta_{\text{RIS2}})$ represents the steering vector corresponding to the expected direction of the direct path between RIS$_2$ and the BS. Thus, if $\hat{x}_j$ is the term that maximizes the Eq. \eqref{sv3}, it is classified as belonging to Path3.}


\vspace{2mm}
\noindent \textbf{Path identification 1:} {Therefore, by estimating the two signals reflected by the RISs, we can predict and identify the direct signal from the UE as the one exhibiting low overall correlation and partial correlation with both previews steering vectors presented. Furthermore, the remaining path can be defined based on the classification of the remaining signal.}

{Figure \ref{fig:AoA} illustrates how the BS functions as the AoA sensor in the proposed system, and how the steering vectors toward the RISs are implemented for comparison with the sensing signals received at the BS.
}

\begin{figure}[H]
    \centering
    \includegraphics[width=0.5\linewidth]{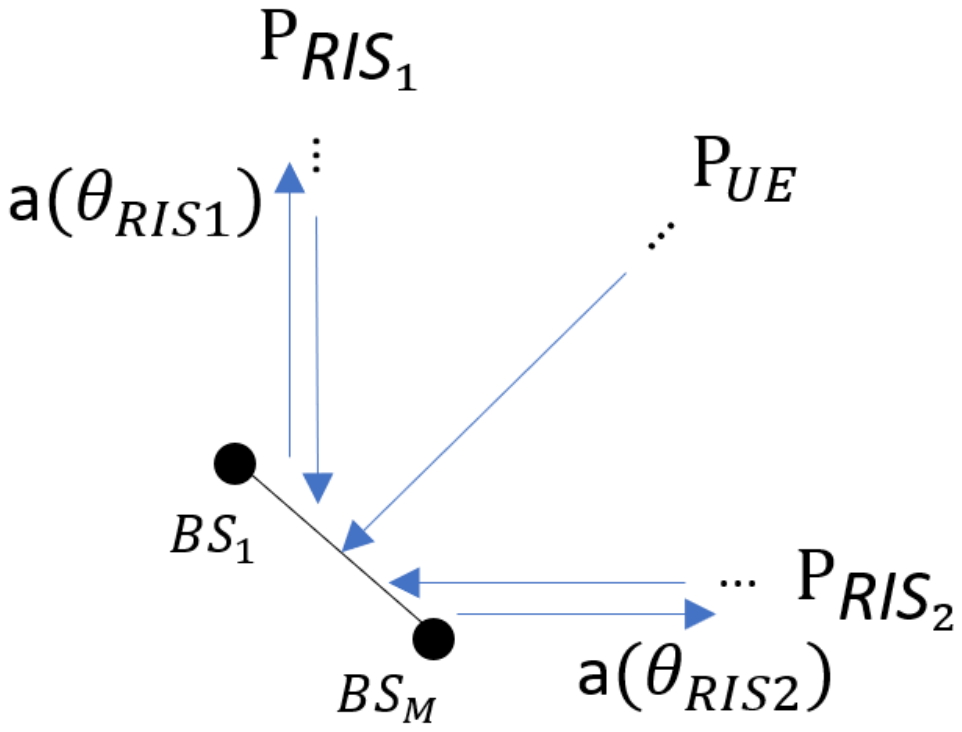}
    \caption{Path identification based on AoA technique.}
    \label{fig:AoA}
\end{figure}


\rf{It is important to highlight that this approach employs a simplified AoA estimation method, aimed not at determining precise angles but at identifying the propagation path necessary for applying the trilateration technique.}

\rf{Alternatively, the MUSIC method provides high accuracy for AoA estimation under well-modeled channel conditions; however, it requires substantial computational resources, particularly in scenarios with a high density of multipath components.}

\subsection{Integration of Communication and Sensing}\label{sec:ics}


\rf{Initially, it is important to emphasize that the RIS is configured only once in a static manner to cover a specific region, typically one with the highest priority. This configuration is performed by the BS through direct communication with the RIS, which programs the RIS elements according to the passive beamforming expression given by \eqref{theta_opt}.}

\rf{The computational complexity of this operation is primarily dictated by the term $\mathcal{O}(MN)$. However, in this initial phase, the system is strategically designed with $M=3$ and $N=3$ to concentrate energy within a wide beam. Under these parameters, the operation's complexity simplifies to a constant, $\mathcal{O}(1)$, resulting in a configuration with exceptionally low numerical complexity.}

To achieve the integration of communication and sensing, the proposed model utilizes time multiplexing for both purposes, optimizing spectral efficiency in modern communication systems.

{The process begins with a sensing transmission from the BS to a specific User ID. In turn, only the UE that identifies the sensing preamble along with the User ID request is able to send a Permission and Response Frame back to the BS. This frame contains preliminary information about the UE’s height temporarily stored. This information is always transmitted to the BS whenever the UE detects the Sensing Frame.}

By applying this user-provided data in conjunction with Eqs. \eqref{eq:placeholder_label1}, \eqref{eq:placeholder_label2}, and \eqref{eq:placeholder_label3}, the BS gathers sufficient information to accurately determine the user's position. \rf{All processing is performed exclusively at the BS, ensuring efficient system operation by eliminating techniques that demand significant processing time and wide bandwidth as observed in other ISAC models, such as inter-BS communication, frequency swapping, beam direction adjustments, and continuous sensing.}

To establish communication, the BS requires a response from the user containing height information after detecting the sensing signal. Once sensing is complete, the BS configures the active beamforming system, dynamically adjusting signal beams toward the user. This enhances sensing accuracy and ensures reliable communication, even in challenging scenarios.

The following steps detail the  {sensing process integrated with communication}:

\begin{enumerate}
    
    \item \textbf{Sensing Signal Transmission:}  The BS emits an encrypted sensing signal using only $3 $ out of $M$ BS antennas, covering a wide area that includes the specific user. This signal is reflected back via the three propagation paths defined in the system model;

    \item \textbf{Call Recognition:}  The user acknowledges the received encrypted sensing signal frame and transmits their height information via permission/response frame, which serves as permission to receive data. This information is critical for this ISAC model to function effectively;

    \item \textbf{Communication Initiation:} Upon receiving the height information and communication permission, the BS utilizes all $M$ antennas to establish communication with the user. Communication continues respecting the cycle period for the sensing, response and communication frames; 

    \item \textbf{Maintaining Directional Communication:}  During communication, the user retransmits their height with permission information whenever the sensing signal frame is detected, ensuring the maintenance of directional communication;

    \item \textbf{Communication Termination:} Communication concludes when the BS stops transmitting the sensing frame. Thus the user to cease sending height information. Alternatively, if the user wishes to end the communication, they can terminate the link by discontinuing the transmission of the link permission frame;

    \item \textbf{End of the Sensing Stage:}  The sensing stage is considered complete when the BS no longer receives packets containing the user's height information.
\end{enumerate}

\subsection{ISAC Complexity Analyses}

{After completing the system framework design, we proceeded to compare the complexity of the ISAC systems presented in the Introduction. Based on the equations and methods described in each paper, we conducted a comparative assessment of the dominant complexities. Then we detailed the complexity of our proposed system to reinforce its potential.} \rf{The objective is to demonstrate that {our proposed system design} maintains low computational complexity, even when incorporating technologies such as RIS.}

\rf{The trilateration-based systems described in \cite{P1} and \cite{P2} exhibit low complexity, as their primary computational burden lies in solving a system of equations to determine the target’s position. The complexity depends on the solution method: Conjugate Gradient results in $\mathcal{O}(n^2)$, while Gaussian Elimination leads to $\mathcal{O}(n^3)$. Given that three-dimensional positioning involves $n=3$ variables, the final complexity simplifies to a constant $\mathcal{O}(1)$, ensuring low computational cost.}

\rf{In contrast, the model in \cite{P3} integrates sensing for multiple targets, increasing overall complexity. The simultaneous detection of $K$ targets requires processing proportional to $\mathcal{O}(K^3)$. Additionally, each target’s localization relies on trilateration, with a complexity of either $\mathcal{O}(n^2)$ or $\mathcal{O}(n^3)$, depending on the solution method. Consequently, the total complexity is given by $\mathcal{O}(K^3 + n^2)$. While computational costs remain low for small $K$, they grow significantly as the number of targets increases.}

\rf{The system analyzed in \cite{P4} employs the MUSIC (Multiple Signal Classification) algorithm, whose dominant complexity is $\mathcal{O}(N^3)$, where $N$ represents the number of RIS elements. Since the RIS system is structured as an $N_X \times N_Y$ matrix, the final complexity becomes $\mathcal{O}((N_X N_Y)^3)$. Additionally, minimizing the autocorrelation matrix $W$ involves a genetic algorithm, though its complexity is not specified in the paper.}

\rf{The model in \cite{P5} introduces complexity related to configuring the Space-Time Coded Metasurface (STCM). The complexity of this process is $\mathcal{O}(NM^3)$, where $M$ denotes the number of BS antennas and $N$ the number of RIS elements. This results in significantly higher complexity compared to the model proposed in this work, as increasing the number of  {RIS elements and BS antennas} leads to higher computational costs.}

\rf{In contrast, our proposed system complexity is primarily dictated by the passive beamforming configuration, as expressed in Eq. \eqref{theta_opt}. The algebraic complexity of this configuration is $\mathcal{O}(MN)$ {in which $M$ denotes the number of BS antennas and $N$ the number of RIS elements}. However, to ensure broad coverage without requiring dynamic reconfiguration, we consider $M=3$ and $N=3$, reducing the complexity to a constant $\mathcal{O}(3 \times 3) = \mathcal{O}(1)$. This guarantees fast convergence and with a low computational cost, as the RIS provides strategic sensing coverage in its static configuration, eliminating the need for continuous reconfiguration. Consequently, the processing time complexity is solely determined by the information signal sharing model, as discussed in Section III.A.}

\section{\rf{Statistical Analysis of Error Probability in an ISAC System}}\label{sec:4}

The proposed formulation introduces a statistical model for the error probability in ISAC systems, accounting for the interdependence between localization accuracy and communication quality.

The error probability in the ISAC system, denoted as $P_E$, arises from two independent events: the localization error and the encrypted sensing communication error, which may occur even when the localization is accurate. Therefore, $P_E$ is defined as,
\begin{equation}
    P_{\text{error}} = P(\text{\{unsuccessful location} \}\cup \text{\{ commun. error}\}).
\end{equation}

The probability of the occurrence of this event can be expressed as,
\begin{align}\label{prob_isac}
    P_{\text{error}}&=P(\text{\{unsuccessful location} \})+P(\text{\{commun. error} \}) \notag\\
    &- P(\text{\{unsuccessful location}\})\cdot P(\text{\{commun. error}\}).
\end{align}

This model enables the prediction of system performance degradation in practical operational scenarios by quantifying the impact of localization accuracy on communication quality. 
In the next subsection, we will adequately quantify the probability of unsuccessful communication, unsuccessful {localization through simulation and discretized by the temporal multiplexing process} and the combination of both unsuccessful probabilities to simulate the overall error probability in the ISAC system.

\subsection{Probability Communication Error}

The probability of unsuccessful communication is traditionally quantified using the bit error rate (BER), i.e. $P(\text{\{commun. error}\})\!\!=\!\! \textnormal{BER}$, a fundamental metric in the performance evaluation of digital communication systems. 
BER measures the fraction of transmitted bits that are incorrectly decoded at the receiver, providing a direct assessment of communication reliability. This parameter is crucial for evaluating the quality of service, as it is inherently linked to the robustness of the modulation scheme and the influence of noise on the transmission channel. In practical scenarios, impairments such as interference, fading, and thermal noise contribute to bit errors. 
The simulated BER, denoted as $\text{BER}_{\text{sim}}$, is statistically defined as the ratio of the total number of erroneous bits to the total number of transmitted bits,
\begin{equation}\label{bersim}
    \text{BER}_{\text{sim}}(t) = \frac{1}{N} \sum_{i=1}^{N} 1(\hat{b}_i \neq b_i),
\end{equation}
in which $N$  represents the total number of transmitted bits in the simulation; $b_i$ denotes the $i$-th originally transmitted bit; $\hat{b}_i$ corresponds to the received bit after demodulation; $1(\hat{b}_i \neq b_i)$ is an indicator function {\cite{indf}} that takes the value $1$ when a bit error occurs ($\hat{b}_i \neq b_i$) and $0$ when the bit is correctly decoded.

The simulated BER is directly influenced by the SNR, as higher noise levels lead to an increased probability of bit detection errors. In {a normalized channel $h$}, the theoretical BER is expressed as a function of the SNR, enabling a comparison between simulation results and analytically expected values. 
For different modulation schemes, the BER-SNR relationship can be determined through closed-form expressions. For instance, in the case of Binary Phase Shift Keying (BPSK) and 16-Quadrature Amplitude Modulation (16-QAM), the BER is given by:
\begin{align}\label{ber_bpsk}
    \text{BER}_{\text{BPSK}}& = \frac{1}{2}Q\left(\sqrt{\frac{{|h|^2} E_b}{N_0}}\right);\notag\\
    \text{BER}_{\text{16-QAM}} &= \frac{3}{8} Q\left(\sqrt{\frac{4 \,{|h|^2} E_b}{N_0}}\right),
\end{align}
where $Q(.)$ express the $Q$-function, a convenient way to represents right tail probabilities for normal random variables \cite{qfunc},  and $\frac{{|h|^2} E_b}{N_0}$  is the bit energy-to-noise power spectral density ratio, which is closely related to the SNR {and normalized by the channel $h$}.

In this system, 16-QAM is used exclusively for sensing applications, utilizing $M = 3$  antennas to transmit encrypted data to a single user, whereas BPSK is adopted for directive communication with $ M = 31$  antennas.

\subsection{Probability Unsuccessful {Localization}}

In this analysis, we assess the probability of unsuccessful localization {in a} scenario with Monte Carlo simulation by varying the success threshold and considering different SNR values.

The methodology is based on a periodic time loop of $0.5$ seconds, as illustrated in Figure \ref{fig:tdd}. At each time instant within this loop, multiple samples are generated using the Monte Carlo method. For each sampling instance, the ToA is recorded, the anchors are identified using the AoA, and with the response frame containing height identification. Consequently, the estimated user position, $\hat{\mathbf{p}}_i$, is determined for the $i$-th sample.

The difference between the actual user position, $\mathbf{p}_i$, in the $i$-th sample and its estimated counterpart is quantified using the Mean Squared Error (MSE), given by,
\begin{equation}
    \text{MSE} = \frac{1}{N} \sum_{i=1}^{N} \left\| \mathbf{p}_i - \hat{\mathbf{p}}_i \right\|^2,
\end{equation}
in which $N$ represents the total number of samples, and $\left\| \mathbf{p}_i - \hat{\mathbf{p}}_i \right\|^2$ denotes the squared Euclidean norm of the error between the real and estimated positions.

The criterion for localization success is constrained by the coverage range of directive communication with the user. Specifically, if the directive communication supports a positioning error threshold ($\epsilon_\text{Lth}$) of up to $1.5$ meters in any direction while maintaining communication quality, then any point within this region is deemed {as} a valid localization.

Accordingly, the simulated {localization} success rate is defined as:
\begin{equation}\label{ssensing}
    \text{SR}_{\text{sim}}^{\text{SENS}} = \frac{1}{N} \sum_{i=1}^{N} 1\left( \left\| \mathbf{p}_i - \hat{\mathbf{p}}_i \right\| \leq \epsilon_\text{Lth} \right),
\end{equation}
in which the indicator function $1(\left\| \mathbf{p}_i - \hat{\mathbf{p}}_i \right\| \leq \epsilon_\text{Lth})$ takes the value of $1$ if the localization error remains within the allowable limit and $0$ otherwise.

Finally, the probability of localization failure is given by the complement of the communication success rate:
\begin{equation}
   P(\text{\{unsuccessful location} \}) = 1 - \text{SR}_{\text{sim}}^{\text{SENS}} .
\end{equation}

\subsection{Simulated Evaluation of ISAC Error}

The error measurement method for the ISAC system, which jointly accounts for localization and communication failures within the simulation framework, follows a formulation analogous to Eq. \eqref{bersim} and \eqref{ssensing}. However, it differs in the counting stage, where the verification condition simultaneously assesses localization accuracy and communication reliability through the indicator function in Eq. \eqref{sisac}:
\begin{equation}\label{sisac}
    \text{SR}_{\text{sim}}^{\text{ISAC}} = \frac{1}{N} \sum_{i=1}^{N} 1\left( \left\| \mathbf{p}_i - \hat{\mathbf{p}}_i \right\| \leq \epsilon_\text{Lth} \land \hat{b}_i = b_i \right),
\end{equation}
here, the indicator function returns 1 only when both sensing and ({$\land$}) communication criteria are satisfied; otherwise, it returns 0. Consequently, this metric quantifies the ISAC system's success rate as the proportion of instances where the localization error remains within thresholds location $\epsilon_{\text{Lth}}$ and the packet is successfully received, i.e., {$\hat{b}_i = b_i$}).

The simulated error probability is derived as the complement of the success rate,  $P(\text{\{ISAC error\}})=1-\text{SR}_{\text{sim}}^{\text{ISAC}}.$
    

%

\section{Numerical Results}\label{sec:5}



Based on the concepts discussed in this paper and the parameters detailed in Table \ref{Tab:parametros}, simulations of the proposed ISAC system were conducted. In all the graphs presented, the coordinate scales are measured in meters, and the color gradients represent variations in SNR gains expressed in decibels (dB).

\begin{table}[htbp]
\centering
\caption{Channel and system parameters used in the simulations.} 
\label{Tab:parametros}

\resizebox{\linewidth}{!}{%
\footnotesize
\begin{tabular}{|c|c|}
\hline
\bf Parameters & \bf Values   \\ \hline
RIS Elements & 3 \\ 
BS Antennas & 31 \\ 
BS Antennas for Sensing & 3 \\ \hline
RIS\(_1\) Position & \begin{tabular}[c]{@{}c@{}} Centralized at (0, 55, 0) \\ with \(\frac{\lambda}{2}\) element spacing\end{tabular}  \\ 
RIS\(_2\) Position & \begin{tabular}[c]{@{}c@{}} Centralized at (55, 0, 0) \\ with \(\frac{\lambda}{2}\) element spacing\end{tabular}  \\ 
BS Position & \begin{tabular}[c]{@{}c@{}} Centralized at (0, 0, 0) \\ with \(\frac{\lambda}{2}\) antenna spacing\end{tabular}  \\ \hline
Beamforming Target (BS) & (25, 25, 0) \\ 
Passive Beamforming Target (RIS) & (-55, 5, 0) \\ 
Transmission Power & 10 W \\ 
Frequency & 10 GHz \\ 
Bandwidth & 500 KHz \\ \hline
\rf{Channel Model} & Rice \\
\rf{Rice factor} & 3\\
\rf{Noise power} & -147dBm\\
{$\epsilon_k$ with $k=1,2,3$} & $\mathcal{N} \left( 0, \frac{|h|^2}{\text{SNR}} \right) $\\ \hline

\rf{Sensing time frame per second} & 40 $\mu$ s\\
\rf{Energy consumption for sensing}  & 0.4 mJ\\ \hline
\end{tabular}}
\end{table}

\subsection{BS and RIS settings}


\rf{One important aspect of this system model is the strategic configuration of the BS and RISs, designed to cover a wide region and enable instantaneous sensing. The resulting performance and associated details are discussed in this section.}

{As a first step, the beamforming pattern is designed to cover the region of interest with sufficient width}. For this purpose, the BS utilizes only $3$ of its $M$ antenna elements, enabling it to form a wide beam. This beam is directed toward an imaginary point at the center of the region, ensuring homogeneous coverage across the entire area.  

The two RISs must also be properly configured to reflect the signals received from the user to reflect back to the BS. A random configuration of the RIS would lead to disordered reflections, forming a hemispherical omnidirectional pattern with low power gain, which is insufficient for system requirements. 
Therefore, it is essential to leverage RIS technology to create an ordered reflection pattern that strategically covers the region, particularly within the first quadrant (1Q), as illustrated in Figure \ref{sismod}. This configuration ensures that the RIS can redirect reflections toward the BS with sufficient power gain.  

To achieve this, the RIS are configured once for the proposed environment, eliminating the need for dynamic reconfigurations during the user detection process. This static approach improves system agility and reduces processing time.  
Simulation tests were conducted, with the BS solely responsible for reconfiguring the RISs based on the passive beamforming equation (Eq. \eqref{theta_opt}).  

{In the first test, it was observed that programming the RIS based on: BS to RIS to cover 1Q, was inadequate. The sensing path passing through the RIS1 (UE to RIS to BS, i.e., Path2 identified in Figure 2) differs from the path which BS used to configure the RIS, because the angle programmed by the BS to RIS to 1Q was replicated for the sensing path in the UE to RIS then reflected to an inappropriate direction.} 

{To address this issue, the RIS elements were reprogrammed based on the proposed path: BS to RIS to cover the second quadrant (2Q). This adjustment ensured the angle is preserved effectively redirecting the sensing signal toward the BS with adequate power gain.}  

Figure \ref{fig:three_subfigures} illustrates the radiation pattern of the system in sensing, specifically:  
a) The radiation patterns emitted by the BS;  
b) The omnidirectional reflection of the signal at the user;  
c) The preconfigured RIS successfully redirecting the sensing signal toward the BS.

\begin{figure*}[t] 
{\includegraphics[width=0.4925\textwidth]{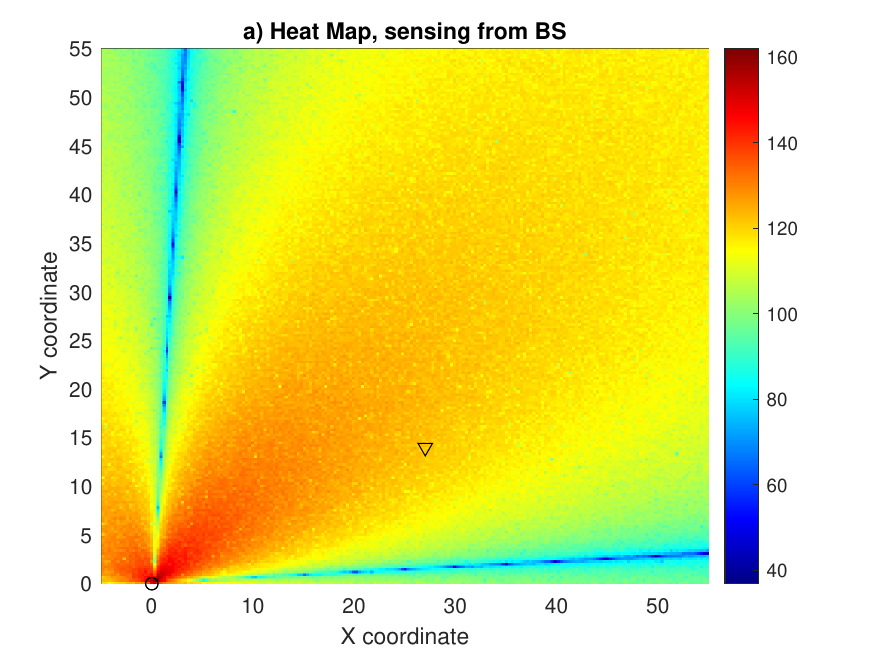}
    \label{fig:subfig1}
}
\hfil 
{\includegraphics[width=0.4925\textwidth]{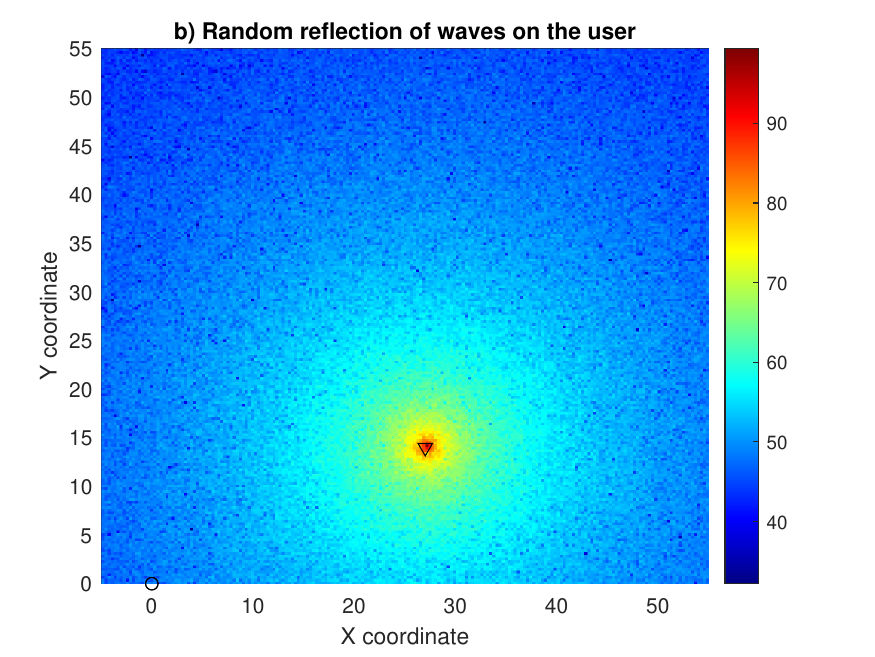} 
    \label{fig:subfig2}
}
\hfil
{\includegraphics[width=0.4925\textwidth]{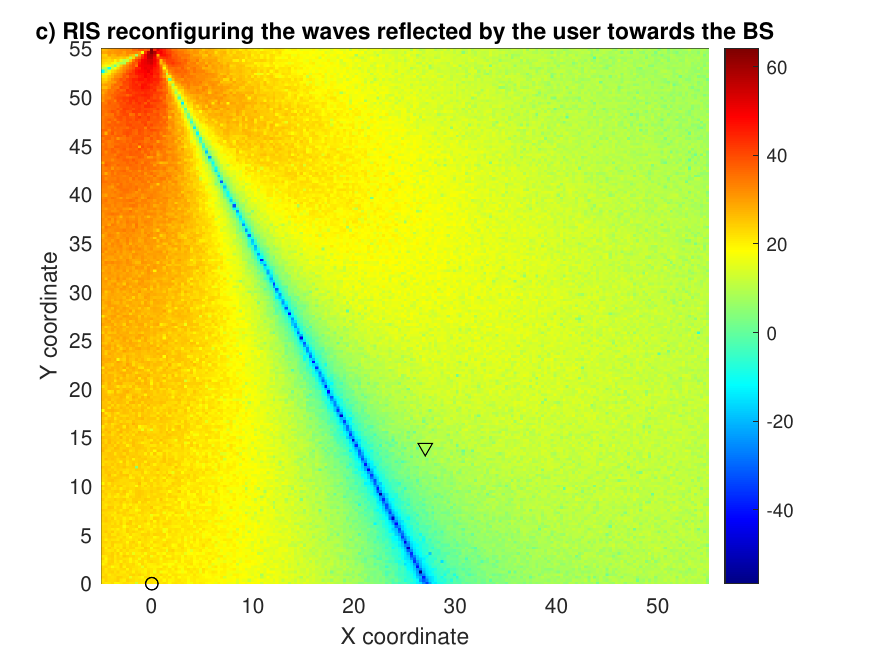} 
    \label{fig:subfig3}
}
\caption{ISAC Sensing, a) BS emission; b) UE reflection; c) RIS-Assisted redirection.}
\label{fig:three_subfigures}
\end{figure*}

\rf{The Figure \ref{fig:commun} illustrates that the data communication beam directed toward the user maintains a minimum beam width of 3 meters in regions closest to the BS. This enables user mobility within a coverage area of $1.5$ meters in any direction during each $0.5$ second cycle, corresponding to the integrated communication and sensing period defined in Figure \ref{fig:tdd}. Consequently, the system ensures reliable communication and sensing for users moving at speeds of up to 10 km/h, which aligns well with typical pedestrian mobility in urban environments. For higher user speeds, the cycle period would require adjustment to maintain performance.}

\begin{figure}
    \centering
    \includegraphics[width=0.8\linewidth]{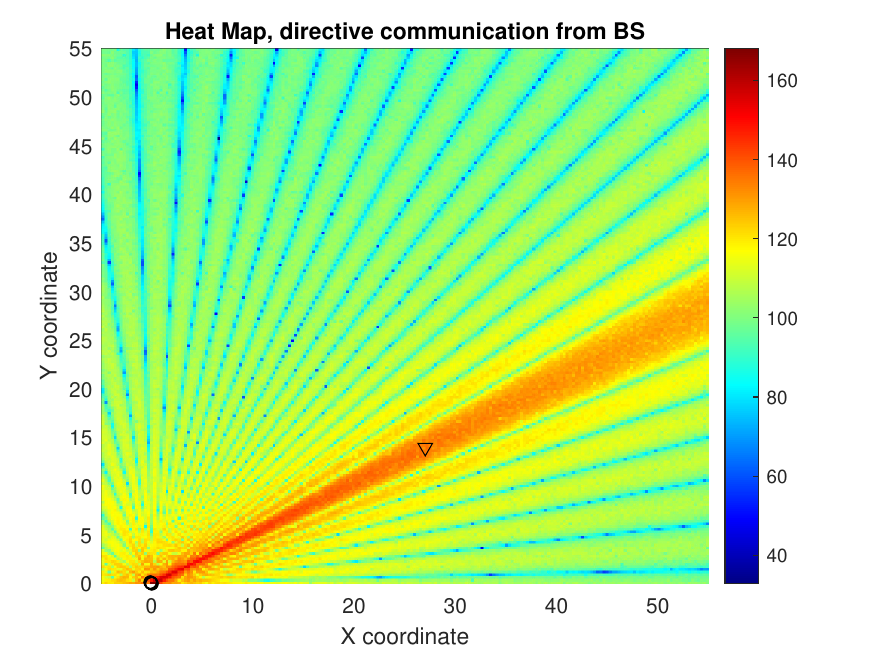}    \caption{Illustration of the ISAC system operating in downlink mode with the BS using 31 antennas.}
    \label{fig:commun}
\end{figure}

In this system, the use of the sensing frame emitted by the BS, combined with specific data for locating a particular user, adopts an approach analogous to the request to send/clear to send (RTS/CTS) method. Specifically, the BS emits a dedicated sensing signal while simultaneously requesting location permission frame from a specific user. If the user consents to being located, they respond by transmitting a frame back to the BS, granting authorization and providing their height. Conversely, if the user does not wish to be located, no response is sent to the BS.

\rf{With the BS emitting the sensing signal at a power of 10 W (40 dBm) over two slots of $20 \mu$s, the total energy consumption for this operation amounts to $0.4$ mJ, equivalent to $1.11 \times 10^{-10}$ kWh. This energy expenditure is extremely low, as expected, given the minimal fraction of time allocated to sensing within the overall system operation.}

\rf{In addition to the energy consumption of the system, it is important to highlight that deploying two static RISs to cover a wide region reduces the requirement for two additional BSs in the trilateration algorithm. Moreover, implementing the technique described in Section \ref{sec:ics}, which provides the user's height, resolves the ambiguity of two potential points in a trilateration system with three anchors, thereby eliminating the need for an additional BS compared to traditional ISAC systems \cite{P2}. This is achieved as the BS identifies the user based on the height information provided.
By combining these two techniques, we effectively eliminate the need for three out of four sensing BSs typically required in the trilateration technique, allowing all processing to be centralized within a single BS.
Notably, the proposed system offers a coverage range five times greater than that of traditional ISAC systems, while maintaining a comparable response time (requiring only a single communication). Compared to ISAC systems employing frequency multiplexing \cite{P5}, the proposed system exhibits half the coverage range but achieves significant improvements in sensing time, operating 45 times faster. Additionally, the proposed approach eliminates the need to allocate and process multiple sensing frequencies, simplifying system operation and reducing complexity.}

\subsection{Defining the ISAC System Coverage Region}

In Figure \ref{fig:visib}, the RIS configuration set to a fixed state, we assessed the areas where the RIS is capable of effectively supporting sensing tasks. In other words, we identified the regions that fall within the visibility range of the BS, showcasing the system’s coverage. The results of this analysis are presented as a heat map in the figure.

\begin{figure}
    \centering
    \includegraphics[width=0.78\linewidth]{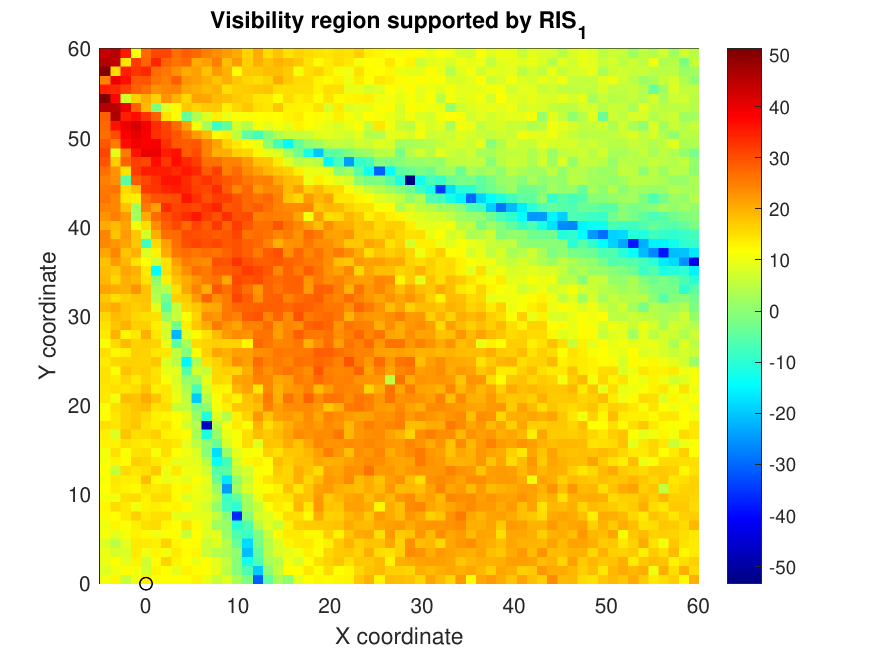}
    \caption{Visibility region provided by the RIS in relation to the BS.}
    \label{fig:visib}
\end{figure}

\rf{In this scenario, each point plotted represents a potential user location, and the accompanying heat map gradient (presented solely for comparison) indicates the signal power received at the BS. Specifically, the signal originates from a reflection in the user’s position, then reflects off the RIS, and finally arrives at the BS. The color scale in the heat map is used to compare how the received power varies across different user locations.}


\subsection{Monte Carlo Simulation}

\rf{Analyzing a single realization, we can observe that the user’s trajectory was recorded at regular time intervals to enable clearer visualization and better understanding of their movement dynamics within the monitored space. Furthermore, we conducted detailed monitoring of a user within the region of distance visibility, analyzing the localization success rate of the models over time.}


{Through $5000$ Monte Carlo simulation trials, the system achieves an accuracy exceeding $98\%$ in determining the user's location within a $1.5$ m radius, even in the presence of timing errors in the ToA ($\epsilon_k$, with $k = 1, 2, 3$) and under the influence of Rice fading}. 


\rf{The error model associated with the signal's return time to the BS was defined based on the heat map results of the visibility region for RIS$_1$ (in Figure \ref{fig:visib}). The variance of the AWGN error model was determined using the middle-case scenario within the visibility region (SNR$=20$dB), as it accounts for areas with SNR values exceeding $40$ dB. For the specific case of $20$ dB, we evaluated the communication success rate for monitoring a user moving within the coverage region. The analysis stipulates that the sensing degree of freedom permits a maximum deviation of $1.5$ m between the estimated and actual positions of the user. This criterion ensures that the user remains within the communication beam, which provides an approximate coverage of $3$ m and is formed by the $31$ BS antennas, as demonstrated in Figure \ref{fig:commun}.}

Additionally, in Figure \ref{fig:ISACsr}.a we evaluate the successful sensing rate as a function of the distance between the user's estimated position and actual position, according to Eq. \eqref{ssensing}, and with different SNR values. This distance is used as a threshold, measured in meters, and represented on the X-axis of the CDF graphs to evaluate the cumulative probability of successful relative to the user's position. 
\rf{ For a $40$ dB SNR, the success rate curves are presented, demonstrating higher success rates and achieving a positioning accuracy of less than $0.5$ m. A mid-range scenario was also analyzed, achieving a positioning accuracy of over $98\%$ within $1.5$ m. Additionally, a scenario with a SNR of $10$ dB was evaluated, which allows for extended coverage regions and increased distances between the BS and the RIS. However, in this {low} SNR scenario, the positioning accuracy decreases to $3$ m, which can significantly degrade communication quality.}

\begin{figure*}[t]
    \centering
    \includegraphics[width=0.4924\linewidth]{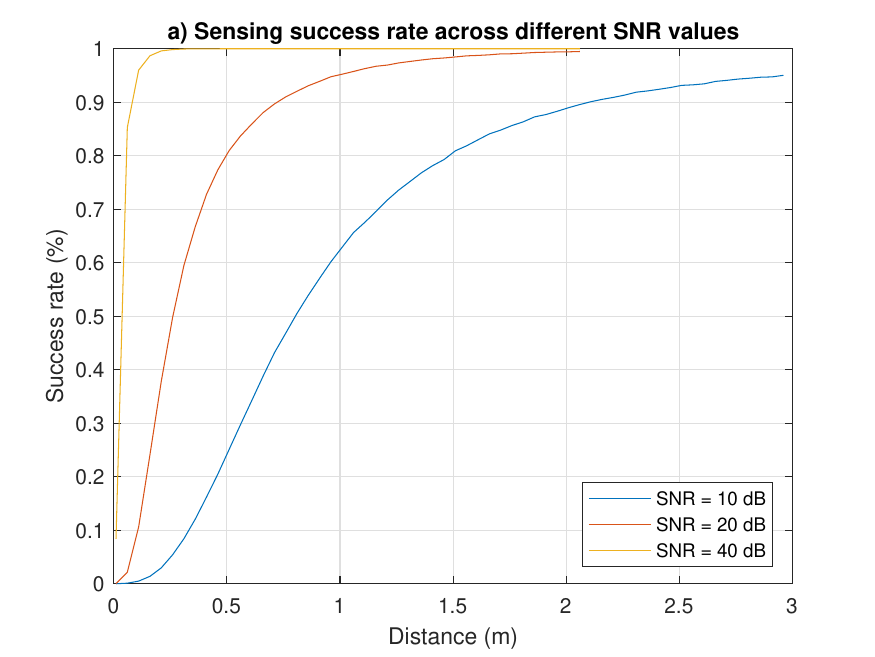}
    %
    \includegraphics[width=0.4924\linewidth]{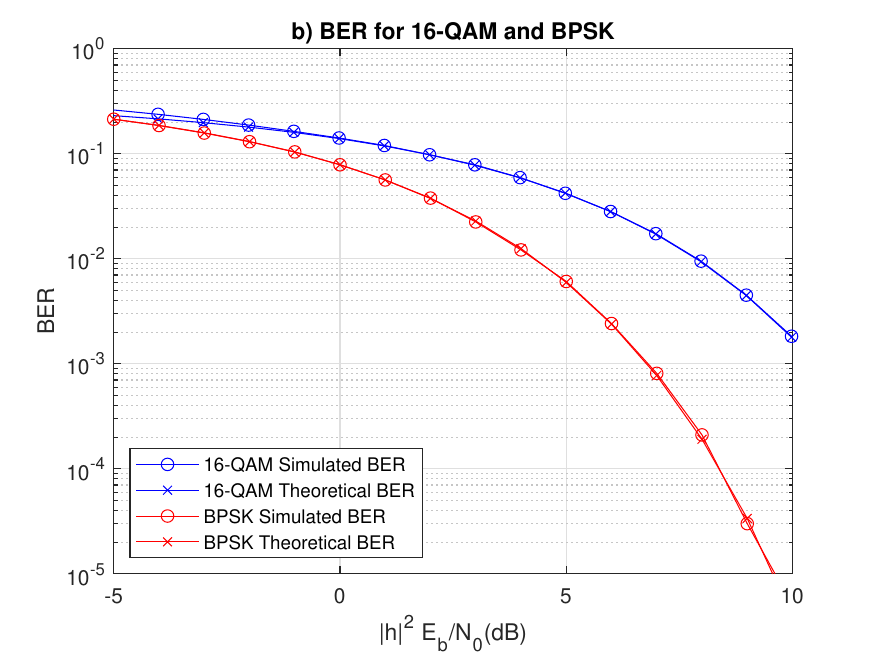}
    \includegraphics[width=0.4924\linewidth]{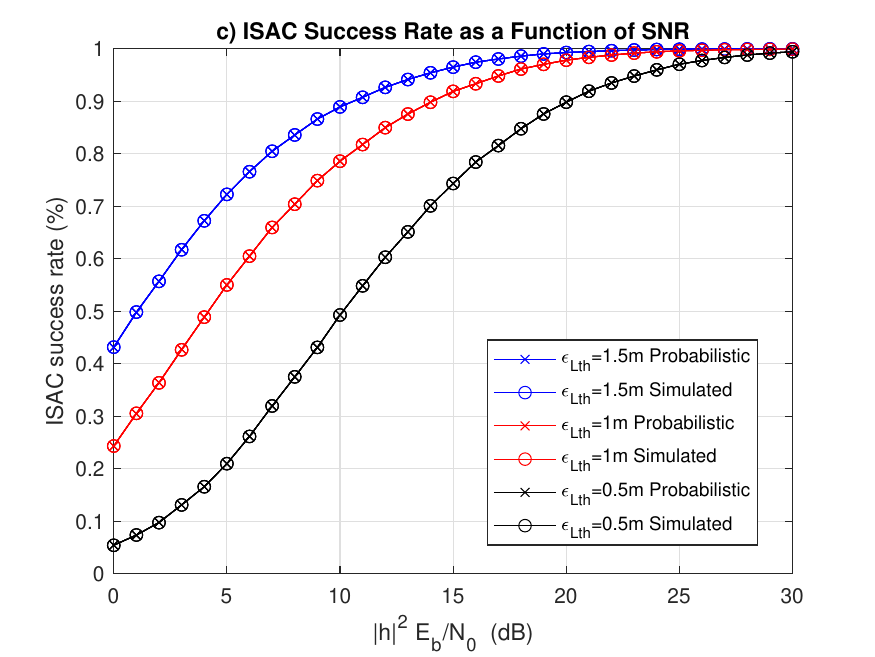}
    
    \caption{a) Impact of SNR on success localization rate precision; b) BER performance as a function of SNR for a normalized channel $h$; c) Impact of SNR on success rate precision for communication based localization.}
    \label{fig:ISACsr}
\end{figure*}




\rf{To validate the ISAC system, it is essential to assess its communication performance. Data transmission quality was evaluated through the BER, using BPSK  modulation, which was selected for its robustness in noisy and fading environments. This modulation ensures efficient transmission even under challenging conditions. However, the proposed system is easily adaptable to alternative modulation schemes (e.g., 16-QAM). Additionally, the modulated signal was transmitted through a channel subject to Rice fading, which represent the typical conditions for wireless communications system. Demodulation at the receiver was performed using a detection technique, ensuring accurate recovery of the transmitted bits.}

    

\rf{Figure \ref{fig:ISACsr}.b shows the variation of BER with respect to SNR, illustrating that, despite natural fluctuations in the channel, the error rate remained low even for an ${{(|h|^2\,E_b)}/N_0}=10\,$dB, reinforcing the reliability of the communication. Over $5\,000$ Monte Carlo realizations for each ${(|h|^2\,E_b)}/N_0$ point, the BER values remained consistently low for both BPSK and 16-QAM, with an average indicating stable and {reliable} communication. {Additionally, the user’s time-varying trajectory under multipath conditions demonstrates that the ISAC system maintains its effectiveness despite spatial variations in the user's position.}}

\rf{This outcome aligns with the high success rate {depicted} Figure \ref{fig:ISACsr}.c whose ISAC success rate are implemented and validated using two different approaches: the simulated and the probabilistic models. The simulation results obtained through Eq. \eqref{sisac}, are represented by curves marked with O, while the probabilistic analytical model, derived from the complement of Eq. \eqref{prob_isac}, i.e., 1 - $P_{\text{error}}$, is represented by curves marked with X.}

    

\rf{Although these models differ, both utilize the location error threshold. The key distinction lies in the formulation of the communication analysis integrated with ISAC. The probabilistic model relies on the theoretical bit error probability for BPSK Eq, \eqref{ber_bpsk}, with the sensing analysis derived from Eq. \eqref{ssensing}, whereas the simulation results are directly implemented using Eq. \eqref{sisac}.}

\rf{Despite being derived through different methodologies, both models produce identical results, validating the integration analysis between the localization and communication systems. Furthermore, a high success rate is achieved for SNR values above 20 dB across all analyzed distance thresholds, suggesting that the dynamic channel variations were effectively mitigated by the characteristics of the ISAC system, thereby enhancing the reliability of communication integrated with sensing.}

Lastly, simulation tests were conducted and revealed that two or more random reflections in cascade eliminate any possibility of creating alternative paths beyond those identified as Path1, Path2, and Path3. This behavior is also observed in the project \cite{P3} validating our multiple random reflections. Moreover, it underscores the system’s robustness. Thus, by applying advanced signal processing techniques, we believe the implementation of multi-user models in this ISAC system is feasible. This prospect will be explored in future studies, extending the system's potential for applications in complex scenarios.

\section{Conclusion}\label{sec:6}

This paper demonstrates the integration of communication and sensing in modern wireless systems through strategical configuration of BS and RIS elements. By employing time multiplexing in ISAC systems, the proposed approach enhances operational performance. Simulations confirm that combining active and passive beamforming with refined designed of BS and RIS configurations is key to effective sensing and communication.
An important characteristic is the use of a static RIS configuration, which covers a wide area and reflects sensing signals back to the BS without requiring dynamic updates or multiple frequency beams, thus reducing processing time. 
The use of trilateration, AoA, and ToA enables accurate localization by combining multipath signals. Integrating these techniques supports effective link selection and minimizes sensing overhead and communication interference.
Furthermore, this paper highlights the significant contribution of properly configuring RISs in ISAC systems to optimize sensing and communication in complex propagation environments. Employing a static configuration tailored to the specific environment, along with advanced beamforming techniques, enhances system agility, reduces processing complexity, and improves sensing accuracy.

\appendices
\section{Converting the Original System into a Linear System}
\label{appendix:proof}

From the original problem presented in Eq. \eqref{Simplf}, the system is reconstructed through algebraic manipulations. Specifically, the quadratic terms are eliminated by subtracting one equation from another. For instance, subtracting Eq. \eqref{eq1} from Eq. \eqref{eq2}, we obtain:
\begin{align}
&d_2^2 - d_1^2=\left[(x - \textnormal{R1}_x)^2 + (y - \textnormal{R1}_y)^2 + (z - \textnormal{R1}_z)^2\right] \notag\\
&- \left[(x - \textnormal{BS}_x)^2 + (y - \textnormal{BS}_y)^2 + (z - \textnormal{BS}_z)^2\right] \notag\\
&{\footnotesize = (x^2\! -\! 2x \,\textnormal{R1}_x\! +\! \textnormal{R1}_x^2 \!+ \!y^2 \!- \!2y\,\textnormal{R1}_y \! +\! \textnormal{R1}_y^2\!+\! z^2\!- \!2z\,\textnormal{R1}_z \!+ \!\textnormal{R1}_z^2)} \notag\\
&{\footnotesize - (x^2 \!- \!2x\,\textnormal{BS}_x \!+ \!\textnormal{BS}_x^2 \!+\! y^2 \!-\! 2y\,\textnormal{BS}_y\!+ \!\textnormal{BS}_y^2\! +\! z^2\! -\! 2z\, \textnormal{BS}_z \!+ \!\textnormal{BS}_z^2). }
\end{align}

Isolating the variables $x$, $y$ and $z$ and simplifying, we obtain,
\begin{align}
    &-2x(\textnormal{R1}_x - \textnormal{BS}_x) - 2y(\textnormal{R1}_y - \textnormal{BS}_y) - 2z(\textnormal{R1}_z - \textnormal{BS}_z) = d_2^2 \notag\\
    & \quad- d_1^2 + \textnormal{BS}_x^2 - \textnormal{R1}_x^2 + \textnormal{BS}_y^2 - \textnormal{R1}_y^2 + \textnormal{BS}_z^2 - \textnormal{R1}_z^2.
\end{align}

This process is repeated to eliminate all quadratic terms involving \(x\), \(y\), and \(z\), resulting in the following system of linear equations:
\begin{align}
    &x(\textnormal{R1}_x - \textnormal{BS}_x) + y(\textnormal{R1}_y - \textnormal{BS}_y) + z(\textnormal{R1}_z - \textnormal{BS}_z) \notag\\
    &= \frac{d_1^2 - d_2^2 \!+\! \textnormal{R1}_x^2 \!-\!\textnormal{BS}_x^2 \!+ \!\textnormal{R1}_y^2 \!-\! \textnormal{BS}_y^2 \!+ \!\textnormal{R1}_z^2 \!- \textnormal{BS}_z^2}{2}\\
    &x(\textnormal{R2}_x - \textnormal{BS}_x) + y(\textnormal{R2}_y - \textnormal{BS}_y) + z(\textnormal{R2}_z - \textnormal{BS}_z) \notag\\
    &= \frac{d_1^2 - d_3^2\! + \!\textnormal{R2}_x^2 \!- \!\textnormal{BS}_x^2\! + \!\textnormal{R2}_y^2\! -\! \textnormal{BS}_y^2 \!+ \!\textnormal{R2}_z^2 \!- \!\textnormal{BS}_z^2}{2}\\
    &x(\textnormal{R2}_x - \textnormal{R1}_x) + y(\textnormal{R2}_y - \textnormal{R1}_y) + z(\textnormal{R2}_z - \textnormal{R1}_z)\notag\\
    &= \frac{d_2^2\! -\! d_3^2 + \textnormal{R2}_x^2 \!- \!\textnormal{R1}_x^2 \!+ \!\textnormal{R2}_y^2 \!-\! \textnormal{R1}_y^2 \!+ \!\textnormal{R2}_z^2\! -\! \textnormal{R1}_z^2}{2}.
\end{align}

This system represents the reformulation of the initial problem into a linear framework and is further simplified through the following assignments, with the results provided in Eqs. \eqref{eq:placeholder_label1}, \eqref{eq:placeholder_label2} and \eqref{eq:placeholder_label3}.

It is important to highlight that this solution aligns with established trilateration models \cite{trilat}. Therefore, its application offers significant advantages in terms of response time and processing efficiency for device localization and sensing.



\end{document}